\documentclass[%
twocolumn,
 amsmath,amssymb,
 aps,
superscriptaddress
]{revtex4-2}

\usepackage{graphicx}% Include figure files
\usepackage{dcolumn}% Align table columns on decimal point
\usepackage{bm}% bold math
\usepackage{multirow}
\usepackage{xcolor}
\usepackage[normalem]{ulem}
\definecolor{newtext}{RGB}{0, 0, 200}
\definecolor{comment}{RGB}{255, 0, 0}
\newcommand{\averaging} [1] {\left \langle \! #1 \!\right\rangle}
\DeclareMathOperator\erfc{erfc}

\begin{document}

\preprint{APS/123-QED}

\title{Selection rules for ultrafast laser excitation and detection\\of spin correlation dynamics in a cubic antiferromagnet}

\author{Anatolii E. Fedianin}
\email{Fedianin.A.E@mail.ioffe.ru}
%\affiliation{ Ioffe Institute, 194021 St. Petersburg, Russia}
\author{Alexandra M. Kalashnikova}
\affiliation{Ioffe Institute, 194021 St. Petersburg, Russia}
\author{Johan H. Mentink}
\affiliation{ Radboud University, Institute of Molecules and Materials,\\
Heyendaalseweg 135, 6525 AJ Nijmegen, The Netherlands}

\date{\today}% It is always \today, today,
             %  but any date may be explicitly specified

\begin{abstract}
Exchange interactions determine the correlations between microscopic spins in magnetic materials. 
Probing the dynamics of these spin correlations on ultrashort length and time scales is, however rather challenging, since it requires simultaneously high spatial and high temporal resolution. 
Recent experimental demonstrations of laser-driven two-magnon modes - zone-edge excitations in antiferromagnets governed by exchange coupling - posed questions about the microscopic nature of the observed spin dynamics, the mechanism underlying its excitation, and their macroscopic manifestation enabling detection.
Here, on the basis of a simple microscopic model, we derive the selection rules for cubic systems that describe the polarization of pump and probe pulses required to excite and detect dynamics of nearest-neighbor spin correlations, and can be employed to isolate such dynamics from other magnetic excitations and magneto-optical effects.
We show that laser-driven spin correlations contribute to optical anisotropy of the antiferromagnet even in the absence of spin-orbit coupling.
In addition, we highlight the role of subleading anisotropy in the spin system and demonstrate that the dynamics of the antiferromagnetic order parameter occurs only in next-to-leading order, determined by the smallness of the magnetic anisotropy as compared to the isotropic exchange interactions in the system. 
We expect that our results will stimulate and support further studies of magnetic correlations on the shortest length and time scale.

\end{abstract}

%\keywords{Suggested keywords}%Use showkeys class option if keyword
                              %display desired
\maketitle

%\tableofcontents

\section{\label{sec:level1}Introduction}

The exchange interaction is the leading interaction in magnetically ordered solids, where the sign and strength dictate the type of magnetic ordering, the temperature of magnetic transitions, and the dispersion of collective magnetic excitations – magnons. 
Ever since the first experiments on ultrafast magnetism, the feasibility of controlling exchange interaction by short laser pulses has been considered as one of the ultimate tasks \cite{MikhaylovskiyNatureComm2015,rhie2003prl,Melnikov2003,JuPRL2004,Thiele2004apl,wall2009prl,forst2011prb,secchi2013annphys,subkhangulov2014,mentink2014prl,MentinkNatureComm2015,matsubara2015ncomms,claassen2017NatComm,Kitamura2017} offering an access to the highest frequencies of spin dynamics with the shortest possible wavevectors \cite{ZhaoPRL2004,BossiniNatComm2016}, and ultimately, to switching material between ferro- and antiferromagnetic states \cite{Li2013nature}. 
This task, however, appeared to be highly challenging. 
The perturbations of the exchange interaction reported so far in experiments appear to be weak and do not exceed a few percent \cite{Bossini2019,MikhaylovskiyNatureComm2015,MikhaylovskiyPRL2020}. 
Thus, it is naturally challenging to isolate this effect from other laser-induced effects, such as opto-magnetic and photomagnetic ones \cite{KimelPhysRep2020}, to establish selection rules for laser-induced exchange perturbation, and to reveal a leading mechanism of light-matter interaction responsible for the effect. 

Another challenge in studying the laser-induced control of short-range exchange interactions stems from the fact that the related spin dynamics may not comply with the classical intuition based on quasi-homogeneous dynamics of magnetic order parameters. So far, different works were describing the spin dynamics triggered by laser-induced perturbation of exchange using terminology originating from quantum information science, such as magnon squeezing \cite{ZhaoPRL2004,ZhaoPRB2006}, and magnon entanglement driving longitudinal oscillations of the antiferromagnetic vector \cite{BossiniNatComm2016,Bossini2019,DeltenrePRB2021}. 
However, as it was emphasized in several theoretical studies \cite{MentinkNatureComm2015,fabiani2022prb,Bossini2019, Kitamura2017,wuhrer2022prb, claassen2017NatComm}, such dynamics is closely related to correlations between spins at nearest-neighbor sites, which therefore is an another important parameter which needs to be analyzed for comprehensive understanding of the spin dynamics triggered by laser-induced exchange perturbation \cite{MentinkNatureComm2015,MentinkJPCM2017,Fabiani2021,DeltenrePRB2021,Bossini2019,fabiani2022prb}. 
Indeed, the sign and magnitude of spin-spin correlations that characterize a particular magnetic state is unambiguously connected with the sign and strength of the exchange interaction. 
However, since spin correlations are not a magnetic order parameter themselves, it remains unclear how such dynamics can manifest itself in optical pump-probe experiments, which measure only macroscopic quasi-homogeneous parameters of the magnetic medium.

To address this problem, it is of particular importance to establish a direct link between the light-matter interaction, microscopic spin correlation dynamics, and its macroscopic manifestation. 
This calls for a theoretical framework, which shows and allows us to model (i) how optical perturbation of exchange interaction excites dynamics of spin correlations, (ii) which macroscopic observables are modulated by the dynamics of spin correlations, and (iii) what is the best strategy to probe this modulation optically.
While the first part of this task is being developed rapidly \cite{MentinkNatureComm2015,MentinkJPCM2017,DeltenrePRB2021,Fabiani2021}, analysis of the rest of the problem is scarce \cite{Kitamura2017} and is limited to consideration of particular materials and signals studied experimentally \cite{ZhaoPRL2004,Bossini2019,Formisano_JPCM2022}.

In this Article we present a theoretical description of laser-driven perturbation of the exchange interaction in an antiferromagnetic dielectric and related detection of the spin dynamics excited by it. 
The perturbation is governed by light-matter interactions in the electric-dipole approximation for the non-dissipative regime. We derive spin correlations for nearest neighbors in case of the simple Heisenberg model for cubic dielectric antiferromagnets with and without magnetic anisotropy. 
We show that dynamics of spin correlations can be excited with ultrashort linear polarized laser pulses yielding an impulsive change of the exchange interaction between nearest neighbors, with a frequency corresponding to the frequency of the two-magnon mode. 
We analyze the effect of the excited spin correlations on macroscopic characteristics of the system, i.e. the dielectric permittivity and the N\'eel vector. 
We show that the laser-induced dynamics of nearest-neighbor spin correlations manifest itself in an induced optical anisotropy. 
Hence the latter can be used as a reliable manifestation of the laser-driven spin correlations dynamics triggered by impulsive perturbation of the exchange interaction, despite that the pairs of magnons involved originate from the edge of the Brillouin zone.
Moreover, the analysis of laser-induced changes of optical anisotropy reveals details of the exchange perturbation, such as the orientation of the bonds along which the exchange interaction was altered.
We further show that longitudinal dynamics of the N\'eel vector can also emerge, however, only in a medium with magnetic anisotropy. 
In contrast to optical anisotropy, the dynamics of the N\'eel vector is found to be less informative regarding the microscopic details of the exchange perturbation. 
Finally, we link our theoretical results to ongoing experimental attempts to observe and comprehend laser-driven two-magnon modes. 
To this end we present complete pump and probe polarization dependencies of modulation of the probe polarization at the frequency of the two-magnon mode, and thus outline the strategy of excitation and detection of this mode in experiments.

The Article is organized as follows. In Sec.~\ref{sec:Theory} we introduce a cubic insulating antiferromagnet as a model system to theoretically analyze the problem of optical excitation and detection of the spin correlations. 
It is followed by formulation of the unperturbed case in terms of the Heisenberg model (Sec.~\ref{sec:Theory:Unperturbed}), establishing a link between Hubbard model describing interaction of light with electronic degrees of freedom and the spin Hamiltonian (Sec.~\ref{sec:Theory:Hubbard}), and obtaining explicit expressions for the modulation of the dielectric permittivity tensor by spin correlations (Sec.~\ref{sec:Theory:Epsilon}).
In Sec.~\ref{sec:Theory:Perturbation} we derive an expression for laser-induced changes of exchange interaction based on the Hubbard model, and in Sec.~\ref{sec:Theory:DynamicsOfSpinCorr} we describe related dynamics of macroscopic medium characteristics.
Sec.~\ref{sec:Results:Oscillations} presents numerical calculations of the laser-induced spin correlation dynamics in model antiferromagnet  KNiF$_3$. 
In Sec.~\ref{sec:Results:Probe} and Sec.~\ref{sec:Results:NeelVector} we show modulation of dielectric permittivity and the N\'eel vector driven by dynamics of spin correlations. 
In Sec.\ref{sec:Results:Anisotropy} we also obtain results for the case of antiferromagnet with additional magnetic anisotropy. 
In Sec.\ref{sec:Discussion} we discuss applicability of the developed theoretical description, and discuss good agreement between our theoretical results and experimental ones reported so far.

\section{\label{sec:Theory}Minimal model for the dynamics of spin correlations in cubic antiferromagnets}

\begin{figure}[t]
\centering{
\includegraphics[width=1\linewidth,keepaspectratio,trim=0in 1px 0in 0px, clip]{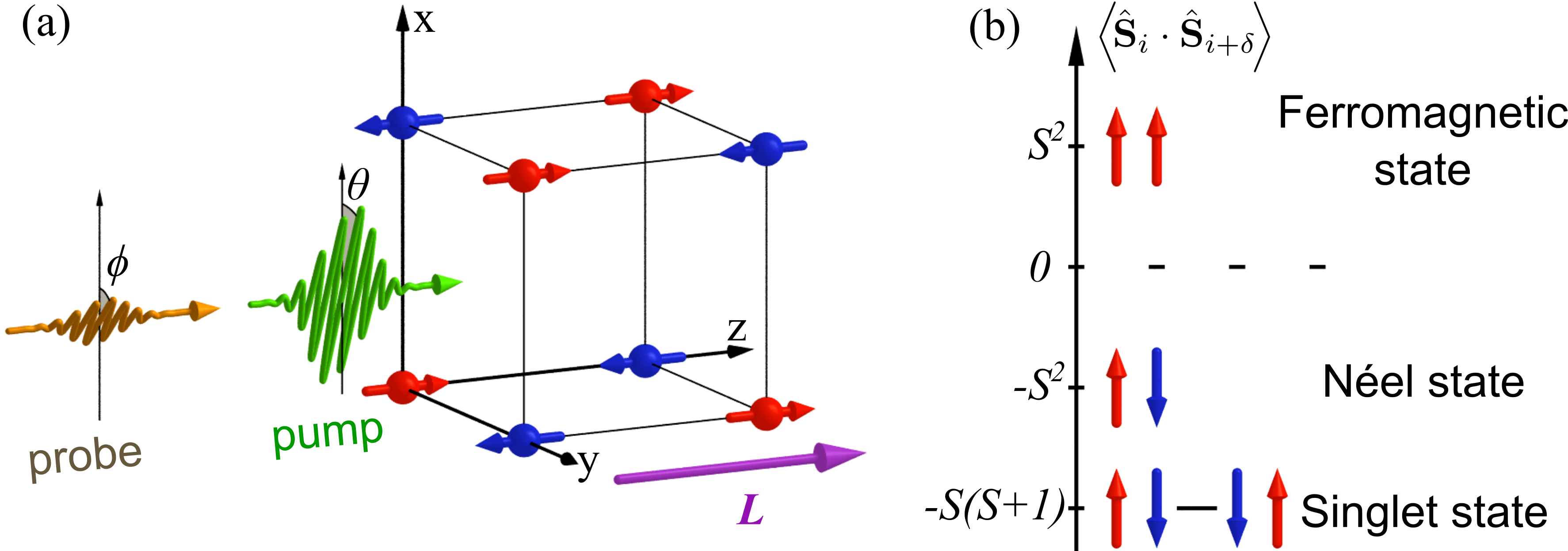}
}
\caption{\label{fig:Setup} 
(a) Illustration of the setup modeled. Magnetic ions form two (red and blue) antiferromagnetically coupled sublattices. $\mathbf{L}$ is the classical antiferromagnetic (N\'{e}el) vector.
Electric field of the incident pump and probe pulses make an angle $\theta$ and $\phi$ with the $x$-axis, respectively.
Pump-induced changes of polarization $\Delta \phi(t)'$ and ellipticity $\Delta \phi(t)''$ of the probe pulses after passing through the sample are the output values.
(b) Schematic diagram illustrating the values of spin correlations in a ferromagnetic, classical antiferromagnetic (N\'{e}el), and singlet antiferromagnetic states.}
\end{figure}

We consider a setup resembling a conventional pump-probe geometry designed to detect laser-induced spin dynamics by optical means (Fig.~\ref{fig:Setup}(a)). 
%The optical pump pulse with duration $2\tau_p$ and polarization angle $\theta$ is perpendicular to one of the crystal axis of a cubic antiferromagnet.
The optical pump pulse with duration $2\tau_p$ and polarization angle $\theta$ is incident along one of the crystal axis of a cubic antiferromagnet.
The spin dynamics excited by the pump pulse is probed by measuring the changes of polarization $\Delta\phi(t)$ of a weaker pulse delayed by the time $t$ with respect to the pump pulse and initially polarized at the angle $\phi$.  
In particular, we consider the pump pulse having Gaussian temporal profile of intensity, and the duration $2\tau_p$ is defined at the $e^{-1}$ level of the peak intensity.

For the medium, we consider the primitive cubic crystal structure with a lattice parameter $a$.
Magnetic ions occupy each corner of a cube.
We consider the G-type antiferromagnetic ordering with each magnetic ion surrounded by the ions with opposite magnetic moment. 
Thus, the magnetic system is described by two \textit{magnetic} sublattices with opposite spin orientations, $S_A$ and $S_B$.
We consider the $z$-axis as quantization axis for the spins, coinciding with the propagation direction of pump and probe pulses, and subsequently generalize to other propagation directions for the pump and probe pulses.
Our focus is mainly on the case of an antiferromagnet possessing no magnetic anisotropy. Correspondingly, we will find no dependence on the actual orientation of spins in space.
Additionally, we analyze how introduction of magnetic anisotropy affects the main results.
Also, where relevant, we include results obtained for a 2D square lattice in the $xy$ plane with the same type of spin arrangement with and without magnetic anisotropy. 
This allows us to evaluate polarization dependencies analytically and illustrate how the dependence on the orientation of the N\'eel vector emerges for models with anisotropy.

Dielectric properties of the lattice are described by strong electron-electron interactions, treating the material as a single band Mott insulator.
We limit our consideration to the case of an antiferromagnet transparent for both pump and probe pulses and treat the interaction of the pulses with the medium in the electric-dipole approximation, which is justified in the visible and near-infrared spectral ranges.
Further, we do not include the effects of the spin-orbit coupling in the analysis of light-matter interaction and exclusively focus on the effect of exchange interactions.

\subsection{\label{sec:Theory:Unperturbed} A ground state properties and static spin correlations} 

We start with the formulation of the unperturbed case. We consider the Heisenberg model described by the Hamiltonian
\begin{align}
    \hat{H}=J\sum_{i \delta} \hat{\mathbf{S}}_{i} \cdot \hat{\mathbf{S}}_{i+\delta},
    \label{eq:Ham:Heisenberg}
\end{align}
where $J$ is an exchange energy of the unperturbed system, $\hat{\mathbf{S}}_i=\hat{\mathbf{S}}(\mathbf{R}_i)$ is a spin operator at a position $\mathbf{R}_i$ in the lattice, and the subscript $\delta$ denotes a vector $\boldsymbol{\delta}=(\delta_x,\delta_y,\delta_z)^T$ to the nearest neighbor.
Following the standard approach, we consider the ground state of the system being close to the N\'{e}el state, and employ truncated Holstein-Primakoff expansion with subsequent Bogolubov transformation \cite{BossiniNatComm2016,Bossini2019} to write the system Hamiltonian in terms of magnon creation and annihilation operators, keeping harmonic contributions only:
\begin{align}
        \hat{H}_{HP}=\sum_{q} \hbar \Omega_q (\hat{\alpha}^{\dagger}_q \hat{\alpha}_q + 
        		\hat{\beta}^{\dagger}_{-q} \hat{\beta_{-q}} + 1),\label{eq:Ham:HP}
\end{align}
where $\hat{\alpha}^\dagger $, $ \hat{\beta}^\dagger $ and $\hat{\alpha} $, $ \hat{\beta} $ are magnon creation and annihilation operators for different magnetic sublattices, respectively, and $\boldsymbol{q}$ is the magnon wavevector, denoted as subscript $q$ for brevity. 
$\Omega_q$ is the magnon frequency found as 
\begin{align}
    \Omega_q =\frac{n_a J S}{\hbar} \sqrt{1-\gamma_q^2};\,\,\,
    \gamma_q=\frac{1}{n_a} \sum_{\delta} e^{i \boldsymbol{q}\cdot \boldsymbol{\delta}},
    \label{eq:Omega}
\end{align}
with $n_a$ being the number of nearest neighbors.
It is convenient to introduce two-magnon $\hat{K}$ operators
\begin{align}
        \hat{K}^z_q &=(\hat{\alpha}^{\dagger}_q \hat{\alpha}_q 
        		+ \hat{\beta}^{\dagger}_{-q} \hat{\beta}_{-q} + 1)/2; \nonumber\\
        \hat{K}^+_q &=\hat{\alpha}^{\dagger}_q \hat{\beta}^{\dagger}_{-q}; \hat{K}^-_q =\hat{\alpha}_q \hat{\beta}_{-q} \label{eq:Koperators},
\end{align}
which act directly on magnon pairs and define number of magnon pairs, as well as their creation and annihilation, respectively. 
Commutation relations for $\hat{K}$ can be obtained from the Bose commutator relations for magnon operators as $\left[\hat{K}^z_q , \hat{K}^{\pm}_q\right]= \pm \hat{K}^{\pm}_q$, $\left[\hat{K}^-_q,\hat{K}^+_q\right] = 2 \hat{K}^z_q$. 

In terms of two-magnon $\hat{K}$ operators the Hamiltonian of the isotropic system reads:
\begin{eqnarray}
        \hat{H}_{2M}=\sum_{q} 2\hbar \Omega_q \hat{K}^z_q. \label{eq:Ham:2M}
\end{eqnarray}
Spin correlations are related to $\hat K$ operators as
\begin{widetext}
\begin{align}
	\averaging{\hat{S}^z_i \hat{S}^z_{i+\delta}} =-S (S+1)+\frac{2 S}{N} 
    		\sum_q  \left( \frac{2 \averaging{\hat{K}^z_q(t)}}{\sqrt{1-\gamma^2_q}}  
             - \frac{\gamma_q \averaging{(\hat{K}^+_q + \hat{K}^-_q)(t)} }
            {\sqrt{1-\gamma^2_q}} \right);\label{eq:SpinCorr:Z}    
\end{align}
\begin{align}
        \averaging{\hat{S}^x_i \hat{S}^x_{i+\delta} +\hat{S}^y_i \hat{S}^y_{i+\delta}} =
        		\frac{2 S}{N} \sum_q  \frac{\cos(\boldsymbol{q}\cdot \boldsymbol{\delta})}{\sqrt{1-\gamma^2_q}} 
        		\left(-2\gamma_q \averaging{\hat{K}^z_q(t)} + \averaging{(\hat{K}^+_{q}+\hat{K}^-_{q}) (t)}\right), \label{eq:SpinCorr:XY} 
 \end{align}
\end{widetext}
where $N$ denotes the number of magnetic ions in the considered volume.
Averaging is done over the unperturbed initial state, hence we considered the Heisenberg picture throughout. Dynamics of the N\'eel vector is directly related to the spin correlations as well, and is given by the relation
\begin{align}
		 \averaging{\hat{L}^z }=\frac{N S}{2} - \frac{1}{n_a S} \sum_{i,\delta}
         		\averaging{\hat{S}^z_i \hat{S}^z_{i+\delta}}.\label{eq:NeelVector}
\end{align}
In terms of classical physics the spin correlations show the mutual orientation for the nearest neighbours spins and the type of ordering, which becomes $S^2$ for pure ferromagnetic and $-S^2$ for the N\'{e}el state.
However, in quantum mechanics the classical N\'{e}el state in antiferromagnets is not the ground state. 
Although long-range N\'{e}el order persists for the 3D Heisenberg model, quantum fluctuations reduce the local spin correlations below the classical limit $-S^2$, with a lower limit $-S(S + 1)$ for the local singlet state of two spins. 
Hence, even though the ground state will be close to the classical N\'{e}el state, the numerical value of the spin correlations is below what is accessible classically, as illustrated in Fig.\,~\ref{fig:Setup}(b).
Spin correlations are strongly dictated by the lattice symmetry, which leads to their equality in ground state for different bonds $\averaging{\hat{\mathbf{S}}_i\cdot \hat{\mathbf{S}}_{i+\delta_x}}=\averaging{\hat{\mathbf{S}}_i\cdot \hat{\mathbf{S}}_{i+\delta_y}}=\averaging{\hat{\mathbf{S}}_i\cdot \hat{\mathbf{S}}_{i+\delta_z}}$.
Dynamics of spin correlations would reveal oscillations of the local correlations with a value that is in between the values of N\'{e}el and local singlet states, and, potentially, for strongly nonlinear dynamics, may extend to the ferromagnetic state.

\subsection{Electro-dipole transitions and spin correlations}\label{sec:Theory:Hubbard} 

In the optical range, light excites electric-dipole transitions, and, therefore, does not pump or probe the spin subsystem directly when spin-orbit coupling is not included. 
Nevertheless, spin-exchange processes do show up even in the absence of spin-orbit coupling. 
A minimal model to describe this is the Hubbard model and we illustrate the link between electric dipole transitions and spin correlations using the Schrieffer-Wolff transformation \cite{schrieffer1966physrev}.

\begin{figure}[b]
\centering{
\includegraphics[width=1\linewidth,keepaspectratio,trim=0in 1px 0in 0px, clip]{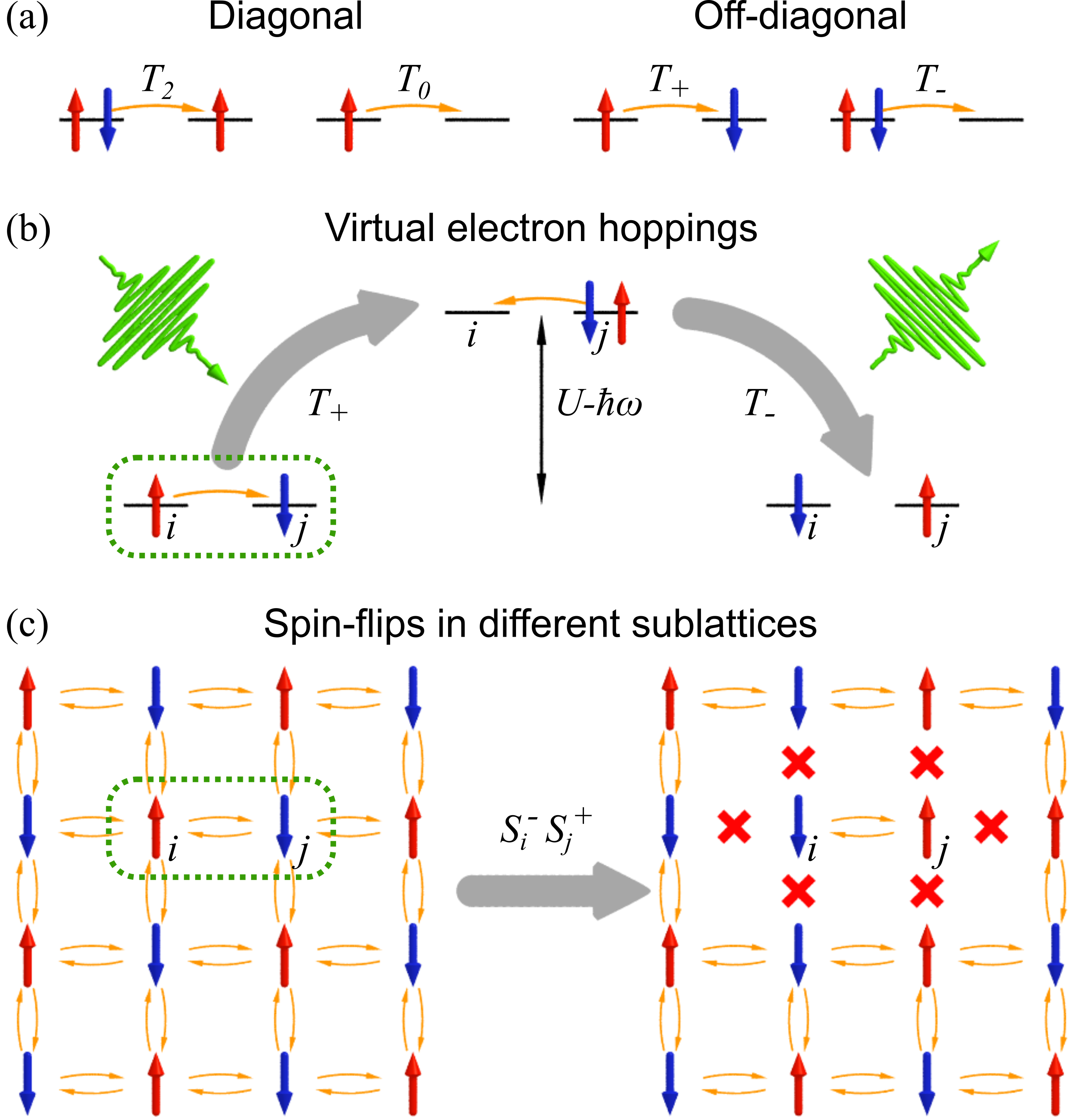}
}
\caption{\label{fig:Hopping} (a) Illustration of diagonal and off-diagonal electron hopping in the Hubbard model, corresponding to $\hat{T}_2$,$\hat{T}_0$ and $\hat{T}_+$,$\hat{T}_-$ .
(b) Mutual electron hopping between two sites excited by a photon with energy $\hbar\omega$. (c-d) Schematic illustration of the pump-induced spin dynamics and the related change of the dielectric response of the 4x4 lattice in the N\'eel state.
(c) Mutual electron hopping between two sites upon the perturbation  of the exchange interaction induced by the pump pulse polarized along the horizontal axis. 
(d) Resulting state of the system with forbidden mutual electron hopping between particular sites. This blockade of the electron hopping influences the coupling of the electric field of the probe pulse with the system, making it different for probe pulses with horizontal and vertical polarization.
}
\end{figure}

We start from a time-independent Hubbard model
\begin{align}
        \hat{H}_{U}&=-\sum_{i j \sigma} \tau_{i j}\hat{c}_{i \sigma}^{\dagger}
        		\hat{c}_{j \sigma}
        	+\frac{1}{2}U\sum_{i} \hat{n}_i \left(\hat{n}_i-1\right), \label{eq:Ham:Hubbard}
\end{align}
where $\tau_{ij}$ is the hopping amplitude between two sites $i$ and $j$, 
$\hat{c}_{i \sigma}$ ($\hat{c}^\dagger_{i \sigma}$) annihilates (creates) an electron at site $i$ with spin projection $\sigma=\uparrow,\downarrow$. $U$ is the on-site repulsion and $\hat{n}_i=\sum_\sigma \hat{c}_{i \sigma}^{\dagger} \hat{c}_{i \sigma}$ is the number operator for electrons at site $i$.
This equation can be decomposed into
\begin{align}
		\hat{H}_{U}&=-\lambda \hat{T} + U \hat{D},\label{eq:Ham:Hubbard_TD}
\end{align}
where $\hat{T}$ is the kinetic energy operator, $\hat{D}$ is the double occupancy operator. 
$\lambda$ stands for the order of hopping, which is numerically equal to 1, but allows one to keep track of the order of expansion in hopping \cite{macdonald1988physrev}. 
The kinetic operator $\hat{T}$ defines hopping between sites in the lattice and can be split into four components schematically illustrated in Fig.~\ref{fig:Hopping}(a). 
Hopping operators $\hat{T}_+$ and $\hat{T}_-$ change
the number of doubly occupied sites and hence the Coulomb energy, while operators $\hat{T}_2$ and $\hat{T}_0$ do not. 
 
As we consider interaction of light with matter in the nondissipative regime, the perturbation of the system is weak and does not change Coulomb energy $U$. 
Therefore, for a further analysis we detach the effects of electrons permutations between sites from changes in double occupancy. 
For that reason we introduce canonical transformation
\begin{align}
	\hat{H}_\mathrm{eff}+U \hat{D}=e^{i \hat{\mathcal{S}}} \hat{H}_{U}  e^{-i \hat{\mathcal{S}}},
    \label{eq:SWTranformation}
\end{align}
with $\hat{\mathcal{S}}$ decomposed by orders of hopping $\hat{\mathcal{S}}=\sum_{i=1}^{\infty} \lambda^{i}$. 
After substitution and following a perturbative expansion we can introduce low energy the effective Hamiltonian
\begin{align}
	\hat{H}_\mathrm{eff}+U \hat{D}&=\hat{H}_{U} +\left[i\lambda 
    		\hat{\mathcal{S}}^{(1)},\hat{H}_{U} \right]\nonumber\\
    	&\quad+\frac{1}{2}\left[i\lambda \hat{\mathcal{S}}^{(1)},
        	\left[i\lambda \hat{\mathcal{S}}^{(1)},
        	\hat{H}_{U} \right]\right] + \mathcal{O}(\lambda^3). \label{eq:Ham:Eff1}
\end{align}
We define $\hat{\mathcal{S}}^{(1)}$ in such a way as to satisfy the equation
$\hat{T}_-+\hat{T}_+=i[\hat{\mathcal{S}}^{(1)},U \hat{D}]$, such all terms changing the number of electrons per site are removed.
Then $\hat{\mathcal{S}}^{(1)}$ can be found as a linear combination of the off-diagonal hopping operators
\begin{align}
	\hat{\mathcal{S}}^{(1)}&=\frac{i }{U} \left(\hat{T}_{+}-\hat{T}_{-}\right). \label{eq:SWOperator}  
\end{align}
By keeping only second order in the hopping expansion we rewrite effective Hamiltonian \eqref{eq:Ham:Eff1} as a combination of the two hopping operators, which acts as a permutation operator, as illustrated in Fig.~\ref{fig:Hopping}(b):
\begin{align}
	\hat{H}_\mathrm{eff}^{(2)}&=-\frac{\lambda^2}{U} \hat{T}_{-}\hat{T}_+, \label{eq:Ham:Eff2TT}
\end{align}
where $\hat{T}_{+}\hat{T}_-$ is neglected due to zero double occupancy of sites in the ground state.

Spin operators can be expressed in terms of creation and annihilation operators and Pauli matrices $2\hat{\mathbf{S}}_{i}=  \sum_{\sigma_1 \sigma_2}\hat{c}^{\dagger}_{i \sigma_1} \boldsymbol{\sigma}_{\sigma_1 \sigma_2} \hat{c}_{i \sigma_2}$. 
Then the effective Hamiltonian \eqref{eq:Ham:Eff2TT} acquires a form
\begin{align}
	\hat{H}_\mathrm{eff}^{(2)}&=\frac{2 \lambda^2}{U} \sum_{i j} |\tau_{i j}|^2  
    		\left(\hat{\mathbf{S}}_{i} \cdot
    		\hat{\mathbf{S}}_{j}-\frac{n_i n_j}{4}\right).
    \label{eq:Ham:Effective}  
\end{align}
Thus, we presented the well-known transformation that connects electron dynamics with spin dynamics. 
In particular, by comparing  Eq.~\eqref{eq:Ham:Heisenberg} and Eq.~\eqref{eq:Ham:Effective}, one sees that 
we obtain the well-known expression $2 |\tau_{i j}|^2/U $ for the exchange interaction.

Hence, as is well known, virtual electron hopping is responsible for the ground state exchange interaction. In the presence of laser-excitation, such virtual hopping processes can be perturbed, leading to optical perturbation of the exchange interactions as we will discuss in the next subsections.

\subsection{\label{sec:Theory:Epsilon}Dielectric permittivity and spin correlations} \label{sec:Theory:Epsilon}

The link between Hubbard model for electric-dipolar transitions and spin correlations can now be used to describe how light-matter interactions change the optical properties and can excite spin dynamics. 
To this end we consider the standard expression for the electric-dipole interaction, which for the Hubbard model can be written as
\begin{align}
    \hat{H}^1(t)&=\frac{a^3}{2} \hat{\mathbf{P}}(t)\cdot\mathbf{E}(t),\label{eq:Ham:Perturbation:Polarization}
\end{align}
where $\hat{\mathbf{P}}=Qa^{-3} \sum_i \hat{n}_i \mathbf{R}_i$ is the polarization operator, with $Q$ the electric charge, and $\mathbf{E}(t)$ is electric field of the pulse.

Next, to rewrite this into an expression that directly reflects the spin correlations, we consider the evaluation of the dielectric susceptibility.
According to the Kubo formulae \cite{Kubo1957}, the dielectric susceptibility and permittivity at a probe frequency $\omega$ are found as

\begin{align}
	\chi^{\mu \nu}(t) &= - \frac{i a^3}{2 \hbar \varepsilon_0}\averaging{\left[ \hat{P}^\mu (t); 
    		\hat{P}^\nu (0)\right]}\mathcal{\theta}(t);\label{eq:nu}\\
  \varepsilon^{\mu \nu}(\omega) &=\varepsilon^{0}I^{\mu \nu } - i \int^{\infty}_{-\infty} \chi^{\mu \nu}(t) 
  		 e^{-i\omega t}  dt,\label{eq:epsilon}
\end{align}
where averaging is over the electronic subsystem, $\hat{P}^{\mu}$ the $\mu$-th component of $\hat{\mathbf{P}}$, $\varepsilon^{0}I^{\mu \nu }$ accounts for all contributions to dielectric permittivity coming from high energy optical transitions, phonons, etc, and $ I^{\mu \nu } $ stands for the identity matrix, $\mu,\nu=\{x,y,z\}$. 

To introduce spin correlations in the expression for the dielectric permittivity (Eq.~\eqref{eq:epsilon}), it is convenient to perform two partial integrations:
\begin{widetext}
\begin{align}
\varepsilon^{\mu \nu}(\omega)
		&=\varepsilon^{0}I^{\mu \nu } - \sum_{n}\frac{a^3}{Z\hbar^2 \omega^2 \varepsilon_0} \langle \psi_n|
        	\left[ \hat{P}^{\mu}, [\hat{P}^{\nu},\hat{H}_U]\right]|\psi_n\rangle \nonumber\\
        &\quad- \sum_{n m} a^3 \frac{e^{-\epsilon_n/k_BT}-e^{-\epsilon_m/k_BT}}
        	{Z\hbar^3 \omega^2 \varepsilon_0} \frac{\langle \psi_n| [\hat{P}^{\mu},\hat{H}_U] |\psi_m\rangle \langle
            \psi_m|[\hat{P}^{\nu},\hat{H}_U] |\psi_n\rangle} {\omega_{n m}+\omega-i 0^+}, \label{eq:varepsilon_W_comm}        
\end{align}
\end{widetext}
where we have formally inserted the full set of electronic eigenstates $\{|{\psi_n}\rangle\}$, with energy levels $\{\epsilon_n\}$. $Z =\sum_n \exp{[-\epsilon_n/k_BT]}$ is the canonical partition sum, with Boltzmann constant $k_B$ and temperature $T$.
As we discussed in Sec.~\ref{sec:Theory:Hubbard}, the nearest neighbor interaction, commutators of the Hubbard Hamiltonian (Eq.~\eqref{eq:Ham:Hubbard}), and polarization operator can be written in a form resembling that of the hopping operator
\begin{align}
\left[\hat{P}^\mu,\hat{H}_{U}\right]
    &=\frac{Q}{a^3}\left[\sum_i \hat{n}_i R^\mu _i,-\lambda\hat{T}\right]\nonumber \\
    &= -\frac{\lambda Q}{a^3} \sum_{i,j} (R^\mu _i-R^\mu _j)\tau_{i,j} \hat{c}^{\dagger}_i \hat{c}_j\nonumber\\
    &= -\frac{\lambda Q}{a^3} \sum_{i,\delta} \delta^\mu \tau_{i,i+\delta} \hat{c}^{\dagger}_i \hat{c}_{i+\delta}.
\end{align}
Hence the commutators yield only kinetic terms, and as we have seen in the previous section (Eq.\,\eqref{eq:Ham:Effective}), we can rewrite them directly in terms of spin correlations. Details of this derivation are given in the Appendix \ref{App:PolToSpin}. 
This yields the main expression for the dielectric susceptibility in terms of the spin correlations
\begin{align}
\varepsilon^{\mu \nu}(\omega) 
        &=\varepsilon^{0}I^{\mu \nu }  + \sum_{\delta} \frac{2 J \left(\averaging{\hat{\mathbf{S}}_{i} \cdot \hat{\mathbf{S}}_{i+\delta}}-\frac{1}{4} \right)}{  U^2-
        	\hbar^2\omega^2} 
            \frac{Q^2 \delta^\mu  \delta^\nu}{\varepsilon_0 a^3},\label{eq:Epsilon:SpinCorr}            
\end{align}
where $\langle \hat{A}\rangle=\sum_{n}Z^{-1}\exp{(-\epsilon_n/kT)} \langle \phi_n| \hat{A}|\phi_n\rangle$, and new wavevectors $|\phi_m\rangle = \exp{(i \lambda\hat{S}^{(1)})} |\psi_m\rangle $. Hence, we have obtained the desired result, which gives us an expression linking the spin correlations along different bonds with the dielectric permittivity. Again, we emphasize that in the ground state $\langle\hat{\mathbf{S}}_i\cdot\hat{\mathbf{S}}_{i+\delta}\rangle$ is the same for all bonds $\boldsymbol{\delta}$, hence the system is optically isotropic in the ground state.

\subsection{\label{sec:Theory:Perturbation}Exchange perturbation} 

With Eq.~\eqref{eq:Epsilon:SpinCorr} we have derived an expression for the dependence of the dielectric permittivity on spin correlations. 
This already suffices for the description of the modulation of the probe pulse due to spin correlations. 
We can use the same expressions as well to describe the interaction of the pump pulse with the antiferromagnet. 
To this end, it is convenient to write the light-matter interaction in a different  way. 
In particular, for relatively small perturbations (in terms of the electric field $\mathbf{E}$ as compared to fields causing dielectric breakdown), we can treat the electrical polarization $\mathbf{P}$ as a linear small perturbation in the electric field $\mathbf{E}$. 
This allows us to use the macroscopic approximation of the linear dielectric susceptibility and define polarization as $\mathbf{P}(\omega)=\chi(\omega)\mathbf{E}(\omega)$.
For the cubic crystals considered here, this means that the light-matter interaction $\hat{H}^1$ can be rewritten as
\begin{align}
    \hat{H}^1&=  \frac{a^3 \varepsilon_0}{2} \sum_{\mu \nu} \int_{-\infty}^\infty \hat{\chi}^{\mu \nu}(t-\tau) E^\mu(t) E^\nu (\tau) d\tau,
\end{align}
where we now use the susceptibility operator $\hat{\chi}$ instead of the expectation value. This change is permitted when working in the Heisenberg representation, where averaging is done with respect to the initial equilibrium state.
            
We first consider monochromatic light $E^\mu(t)=\frac{1}{2}(E^\mu_0 e^{i\omega_p t}+E^{\mu *}_0 e^{-i\omega_p t})$, and evaluate the convolution using the Fourier transform:
\begin{align}
     \hat{H}^1
        &= \frac{a^3 \varepsilon_0}{2}\sum_{\mu \nu} \int_{-\infty}^\infty \hat{\chi}^{\mu \nu}(\omega) e^{i \omega t} E^\mu(t) E^\nu(\omega) d\omega, \label{eq:Ham:Perturbation:Susceptibility}
\end{align}
where
\begin{align}
    \hat{\chi}^{\mu \nu} (\omega)
        &= \sum_{\delta} \frac{2 J \left(\hat{\mathbf{S}}_{i} \cdot \hat{\mathbf{S}}_{i+\delta}-\frac{1}{4} \right)}{ U^2-\hbar^2\omega^2} \frac{Q^2 \delta^\mu  \delta^\nu}{\varepsilon_0 a^3}.
\end{align}
The resulting integral yields both a static and oscillatory contributions. 
Since we are interested in the low-frequency spin dynamics ($\omega_p\gg\Omega_{2M}$), we can average over fast oscillations of the optical pulse, which yields 
\begin{align}
    \hat{H}^1
     &= \frac{a^3 \varepsilon_0}{2}\sum_\mu \frac{1}{2}\hat{\chi}^{\mu \nu}(\omega_p) E^\mu_0 E_0^\nu.
\end{align}
This allows us to write $\hat{H}^1$ as a perturbation of the spin subsystem alone: 
\begin{align}
    \Delta \hat{H}_0
     &=\Delta J \sum_{i, \delta } (\boldsymbol{e}\cdot\boldsymbol{\delta}/a)^2 
             \left(\hat{\mathbf{S}}_{i} \cdot \hat{\mathbf{S}}_{i+\delta}-\frac{1}{4} \right),
\end{align}
where $\boldsymbol{e}=\mathbf{E}_0/|\mathbf{E}_0|$ is the polarization vector of the electric field, and
\begin{align}
    \Delta J &=\frac{J E^2_0 Q^2 a^2}{2  \left(U^2-\hbar^2\omega_p^2\right)}.\label{eq:DeltaJ:Gauss}
\end{align}

Eq.~\eqref{eq:DeltaJ:Gauss} describes the response of the system to monochromatic laser excitation.
To extend this result to the Gaussian-like pulses $E(t)=E_0 \exp{(-t^2/2\tau_p^2)}\cos(\omega_p t)$, we pay attention to the fact that the considered pump frequencies are far away from optical transitions, and one can neglect the spectral dispersion of the dielectric susceptibility, so $\hat\chi(\omega)$ can be replaced with constant value $\hat\chi(\omega_p)$.
Then the perturbation acquires time-dependence dictated by the temporal profile of the pump pulse:
\begin{align}
    \Delta \hat{H}_0
     &= f(t) \Delta J \sum_{i, \delta }  (\boldsymbol{e}\cdot\boldsymbol{\delta}/a)^2 
             \left(\hat{\mathbf{S}}_{i} \cdot \hat{\mathbf{S}}_{i+\delta}-\frac{1}{4} \right),\label{eq:Ham:Perturbation:Spin}
\end{align}
where $f(t)=\exp(-t^2/ \tau_p^2)$. 
We note that $f(t)$ is not limited to a Gaussian pulse shape, but can describe an arbitrary pulse shape with the spectral width not broad enough to require taking into account the dispersion of the dielectric susceptibility within the pulse spectrum.

Thus, in this section we have derived the key expressions which enable one to describe consistently the excitation of spin system by a laser pump pulse (Eq.~\eqref{eq:Ham:Perturbation:Spin}) and the modulation of the dielectric function (Eq.~\eqref{eq:Epsilon:SpinCorr}).
We note that the results are identical to the one obtained directly from expanding the effective Hamiltonian to leading order in the electric field \cite{MentinkNatureComm2015}.
In the following section we apply these expressions to describe the major features of the laser-driven spin dynamics that appears in this description as well as the corresponding changes of optical properties.

\subsection{Dynamics of spin correlations} \label{sec:Theory:DynamicsOfSpinCorr}

In Sec.~\ref{sec:Theory:Perturbation} the light-matter interaction is described in terms of perturbation of exchange interactions, which induces an asymmetry in the system since bonds along different crystal axes are perturbed differently depending on the polarization $\boldsymbol{e}$ of the electric field. We now evaluate the spin dynamics which is triggered by such a perturbation. Since akin to the unperturbed Hamiltonian (Eq.~\eqref{eq:Ham:Heisenberg}), Eq.~\eqref{eq:Ham:Perturbation:Spin} is defined in terms of exchange terms, we can express it in magnon-pair operators. Substitution of the transformation used in Sec.~\ref{sec:Theory:Unperturbed}, we obtain for $\Delta \hat{H}_0$ the following two-magnon description: 

\begin{equation}
    \begin{gathered}
        \Delta \hat{H}_{2M}=f(t) \sum_{q} \left[\hbar \Delta \Omega_q \hat{K}^z_q 
        		+ \hbar V_q \left(\hat{K}^+_q+\hat{K}^-_q\right)\right],
    \end{gathered}\label{eq:Ham:Perturbation:2M}
\end{equation}
with
\begin{align}
        \Delta\Omega_q &=\zeta \Delta J S / \hbar \frac{1-\xi_q \gamma_q}{\sqrt{1-\gamma_q^2}};\\
        V_q &=\zeta \Delta J S / \hbar \frac{\xi_q - \gamma_q}{\sqrt{1-\gamma_q^2}},\label{eq:Vq}
\end{align}
where $\zeta=\sum_{\delta} (\boldsymbol{e}\cdot \boldsymbol{\delta}/a)^2 $, $\xi_q=1/ \zeta \sum_{\delta} (\boldsymbol{e}\cdot \boldsymbol{\delta}/a)^2 \exp (i \boldsymbol{q} \cdot\boldsymbol{\delta})$. Below we consider ultrashort pulses for which $\tau_p\Omega_q \ll 1$. In addition, typically $\Delta J \ll J$, hence it follows that $\Delta \Omega_q \ll \Omega_q$. The change of the magnon frequency is a small perturbation, which is only present during the pulse, leading to a phase shift that is of next order in smallness: $\tau_p\Delta\Omega_q \ll \tau_p\Omega_q\ll 1$, which we ignore in further discussions. 

To evaluate the spin dynamics triggered by the optical perturbation of exchange, we keep only leading order of the two-magnon operators and omit the dynamics for times $t<\tau_p$ during the pulse. 
 The actual derivations are presented in Appendix \ref{App:Kdynamics} and are obtained by using the Kubo formula for linear-response dynamics and by employing the Green-function method for the dynamics in response to a perturbation with a Gaussian profile.
The resulting dynamics for the expectation values for the $\hat{K}$ operators are:

    \begin{align}
        \left\langle\! \hat{K}^z_q(t) \!\right\rangle
        	&=\left\langle\! \hat{K}^z_{q,0}\!\right\rangle;\nonumber\\
        \left\langle (\hat{K}^+_q + \hat{K}^-_q)(t)\right\rangle 
        	&=-4 V_q \sigma_q \left\langle \hat{K}^z_{q,0}\right\rangle \sin{(2 \Omega_q t)}; \nonumber\\    
        \left\langle (\hat{K}^+_q-\hat{K}^-_q)(t)\right\rangle
        	&=-4 i   V_q \sigma_q  \left\langle\! \hat{K}^z_{q,0}\!\right\rangle \cos (2\Omega_q t).
\end{align}
Here we used that expectation values $\averaging{\hat{K}^x}$ and $\averaging{\hat{K}^y}$ are zero in the ground state, and
\begin{align}
\sigma_q=\sqrt{\pi} \tau_p e^{-\Omega_q^2 \tau_p^2} \label{eq:sigma}
\end{align}
is the weighting factor accounting for the Gaussian pump pulse duration.

Combining these expressions with Eqs.~(\ref{eq:SpinCorr:Z},\ref{eq:SpinCorr:XY}) we obtain the analytical expressions describing the time-dependent spin correlations as a result of laser excitation:
\begin{widetext}
\begin{align}
	\averaging{\hat{{S}}^z_i(t) \hat{{S}}^z_{i+\delta}(t)} =-S(S+1)+\frac{4 S}{N} 
    		\sum_q \frac{\averaging{\hat{K}^z_{q,0}}}{\sqrt{1-\gamma^2_q}}  \left( 1 
             + 2\gamma_q V_q \sigma_q \sin{(2 \Omega_q t)} \right); \label{eq:SpinCorr:DynamicsZ}   \\    
        \averaging{\hat{{S}}^x_i(t)  \hat{{S}}^x_{i+\delta}(t) +\hat{{S}}^y_i(t) \hat{{S}}^y_{i+\delta}(t)} =
        		\frac{4 S}{N} \sum_q  \frac{\averaging{\hat{K}^z_{q,0}}\cos(\boldsymbol{q}\cdot \boldsymbol{\delta})}{\sqrt{1-\gamma^2_q}} 
        		\left(-\gamma_q - 2 V_q \sigma_q \sin{(2 \Omega_q t)}\right) ;\label{eq:SpinCorr:DynamicsXY}\\
        \Delta\!\averaging{\hat{\mathbf{S}}_i(t) \cdot \hat{\mathbf{S}}_{i+\delta}(t)} =\Delta J \frac{8 S^2 \zeta }{N \hbar} 
    		\sum_q \frac{(\gamma_q - \cos(\boldsymbol{q} \cdot\boldsymbol{\delta})) (\xi_q-\gamma_q) }{1-\gamma^2_q}   \averaging{\hat{K}^z_{q,0}} \sigma_q\sin (2 \Omega_q t). \label{eq:SpinCorr:DynamicsDelta}
\end{align}
\end{widetext}

From Eqs.~(\ref{eq:SpinCorr:DynamicsZ},\ref{eq:SpinCorr:DynamicsXY}) one sees that the excitation of the system results in spin correlations becoming time-dependent and oscillating at the frequency $2\Omega_q$. 
Notably, the amplitude of these oscillations is defined, in particular, by the parameter $V_q$ (Eq.~\eqref{eq:Vq}), which depends on $\xi_q$ and therefore on the pump pulse polarization angle $\theta$. 

\begin{figure}[t]
\centering{
\includegraphics[width=1\linewidth,keepaspectratio,trim=0in 1px 0in 0px, clip]{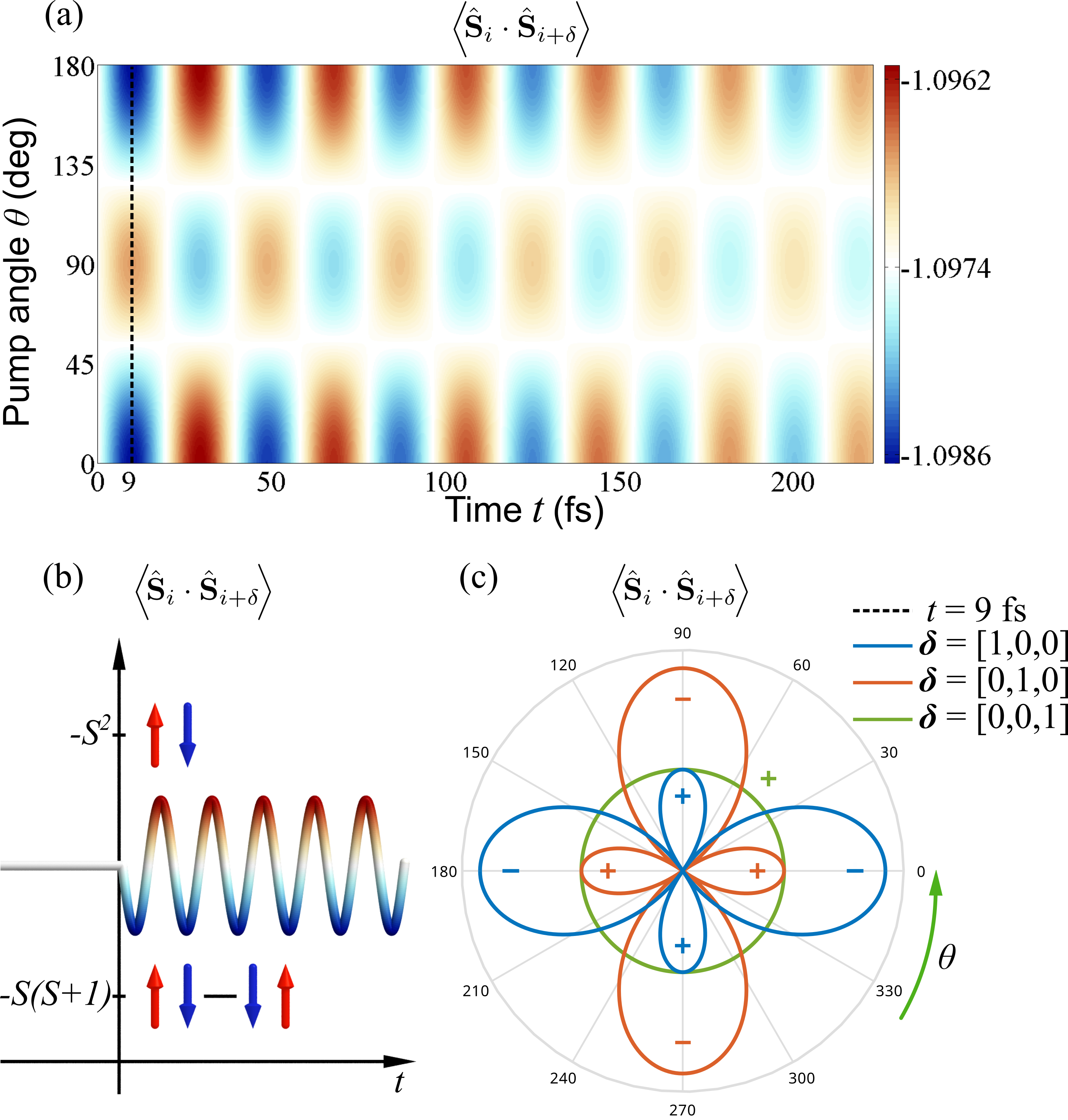}
}
\caption{\label{fig:SS} (a) Dynamics of the spin correlation between nearest neighbours along $x$-axis with respect to the polarization angle of the pump pulse $ \mathcal{\theta} $. Calculations are performed for taking exchange interaction $J=$8.8 meV, temperature $T=0$, laser-induced perturbation $\Delta J=0.008 J$, and the pump duration $2\tau_p$=10\,fs.
(b) Schematic picture of oscillations on the scale of the spin correlations.
(c) Spin correlations at $t=9$\,fs after pulse as a function of the the polarization angle of the pump pulse $\theta$ for nearest neighbours along $x$-, $y$-, and $z$- bonds. $+/-$ indicates increase or decrease of the value with respect to the equilibrium one.}\label{fig:SpinCorr}
\end{figure}

We note also that the time-dependent part of the spin correlations is dependent on $\sigma_q$ (Eq.~\eqref{eq:sigma}) which shows that the amplitude of the spin correlations dynamics diminishes for long pump pulses ($\Omega_q\tau_p\gg1$). This is in agreement with well-established role of the pump pulse duration in impulsive excitation of various coherent processes \cite{ImasakaPRB2018}.
In addition, the correlation dynamics depends on the bond defined by $\boldsymbol{\delta}$. In particular, correlations along different bonds delta are not independent, since for the dynamics at $t>\tau_p$, $\sum_\delta \Delta \averaging{\hat{\mathbf{S}}_i \cdot \hat{\mathbf{S}}_{i+\delta}} = 0$; the total energy is conserved after the interaction of the system with the pump pulse.
Together the results of Secs.~\ref{sec:Theory:Epsilon},\ref{sec:Theory:DynamicsOfSpinCorr}, can describe both excitation and detection of spin correlation dynamics within a single framework.

\section{Results}\label{sec:Results}

In this section, we show the dependencies of dynamics of (i) spin correlations (ii) N\'{e}el vector, (iii) probe modulation on the polarization angle $\theta$, and (iv) influence of magnetic anisotropy on dynamics of the N\'{e}el vector. We start with the ideal cubic lattice and subsequently focus on the anisotropic case. 
All results are based on numerical evaluation of Eqs.~(\ref{eq:Epsilon:SpinCorr},\ref{eq:SpinCorr:DynamicsDelta}), using parameters taken for the prototype KNiF$_3$, given in Table \ref{table:parameters}, and for a cubic system of dimension $V=(L\cdot a)^3$, where $L=30$ is the number of sites in one direction, which is chosen such that convergence is achieved for all observables presented.

\subsection{Oscillations of spin correlations}\label{sec:Results:Oscillations} 
In order to illustrate the dependence of laser-driven spin correlation dynamics on the time $t$ and pump polarization $\theta$, we plot in Fig.\,\ref{fig:SpinCorr}(a) $\averaging{\hat{\mathbf{S}}_i(t) \cdot \hat{\mathbf{S}}_{i+\delta_x}(t)}$ for two nearest neighbours along the $x$-axis.
The duration of the pump pulse was $2\tau_p=10$~fs, and the induced change of the exchange interaction was $\Delta J/J=0.008$.
The feasibility of this value of $\Delta J/J$ is discussed below in Sec.\,\ref{sec:Experiment_values}.
As seen from the calculated dependencies of $\averaging{\hat{\mathbf{S}}_i(t) \cdot \hat{\mathbf{S}}_{i+\delta_x}(t)}$ on time $t$ after excitation and the pump polarization $\theta$, the pump pulse incident along $z$-axis and polarized along either of the $x$- or $y$-bonds excites oscillations of the spin correlations. 
%The frequency of the oscillations is dominantly determined by that of the magnons at the edge of the Brillouin zone, $2\Omega_{q^*} / 2 \pi$= 25\,THz, where $\boldsymbol{q}^*=(\pm\pi/a,0,0)$, since these magnon pairs provide the largest contribution to the nearest-neighbor spin correlations.  
The frequency of the oscillation is $2\Omega_{q^*} / 2 \pi$= 25\,THz and dominantly determined by magnons at the edges of the Brillouin zone: $\Omega(\boldsymbol{q}^*)=\Omega(q^*,0,0)=\Omega(0,q^*,0)=\Omega(0,0,q^*)$, with $q^*=\pm\pi/a$, since these magnon pairs provide the largest contribution to the nearest-neighbor spin correlations. Apparent damping of the oscillations is a result of two-magnon excitation with lower wavevectors.

\begin{table}
    \centering
\caption{Material parameters of KNiF$_3$ used in calculations}
\begin{tabular}{c|c|c|c|c}
  \hline
  % after \\: \hline or \cline{col1-col2} \cline{col3-col4} ...
   $J$ & $S$ & $U$ & $a$ & $\varepsilon^{xx}$\\
   meV &  & eV & \AA & \\
   \hline
   8.8 & 1 & 6.2 & 4 & 2.14 \\
  \hline
  \cite{Pisarev1974} &\cite{lines1967}& \cite{Mattheiss1972} & \cite{deJongh_PhysB1975} & \cite{Dubrovin2021} \\
  \hline
  \hline
 
\end{tabular}
    \label{table:parameters}
\end{table}

Figure\,\ref{fig:SpinCorr}(b) schematically illustrates that the relation between the sign of the perturbation of exchange and the initial phase of the spin correlations. An instantaneous increase of the exchange interaction between spins along $x$-axis induced by the pump pulse polarized along the same axis ($\theta=0^\mathrm{o}$), excites dynamics of spin correlations along this $x$-axis with an initial phase that initially evolve closer the singlet state. Conversely, as shown in Fig.\,\ref{fig:SpinCorr}(a)), a pump pulse along the $y$-axis ($\theta=90^\mathrm{o}$), perturbing only exchange bonds along the $y$-axis, changes the initial phase of the $x-$correlations, driving them initially closer to the N\'eel state.

Even on the basis of energy conversation, not all phases of nearest-neighbor correlations can be identical, since their sum should be conserved during the dynamics after the pump pulse. 
For the 2D case, it was shown that $x$ and $y$ correlations have opposite sign \cite{Fabiani2021} due to symmetry of the lattice and the light-matter interaction. From our numerical results we see that the same asymmetry in phase is present in 3D as well, but amplitudes are not equal anymore.
In Fig.\,\ref{fig:SpinCorr}(c) we summarize how the initial phase and amplitude of the spin correlation between the spins along all three bonds depends on the pump polarization $\theta$. Here, $\Delta\averaging{\hat{\mathbf{S}}_i(t) \cdot \hat{\mathbf{S}}_{i+\delta_\nu}(t)}$ %, $\nu=x,y,z$ 
is plotted at $t=$9\,fs, for which the first maximum of the oscillation occurs (see vertical line in Fig.\,\ref{fig:SpinCorr}(a)). We emphasize that 
perturbation of the exchange interaction along the $x$-bonds results in oscillations of spin correlations along all three bonds, with amplitude being the largest for the $x$-bond, and with opposite phase and twice as small amplitude for $y$- and $z$-bonds. 
For general $\theta$ we find that the pump polarization dependencies of the spin correlations for $x$- and $y$-bonds possesses a two fold symmetry, while the one for the $z$-bond is isotropic. 

\subsection{Dynamics of the N\'{e}el vector}\label{sec:Results:NeelVector}

In the previous section it was shown that the sum of all correlations, $\sum_{\delta}\langle\hat{\mathbf{S}}_i\cdot\hat{\mathbf{S}}_{i+\delta}\rangle$, is time-independent, consistent with total energy conservation. This is true not only for the scalar product, but also for $z$-components of the correlations: $\sum_{\delta}\langle\hat{S}_i^z\hat {S}^z_{i+\delta}\rangle$. Hence, the macroscopic characteristic of long-range magnetic order, the N\'{e}el vector defined by Eq.~\eqref{eq:NeelVector}, will be time-independent.
As it has been argued in \cite{BossiniNatComm2016,Bossini2019}, it is expected that the excited two magnon mode should manifest itself in longitudinal dynamics of $\left\langle L^z\right\rangle$.
However, our numerical calculations show that oscillations of $\left\langle L^z\right\rangle$ are absent under perturbations of exchange interactions, in accordance with the fact that both $\hat{H}$ and the light-matter interaction $\Delta\hat{H}$ are invariant under spin rotations.
Likewise, $\left\langle L^x \right\rangle$ and $\left\langle L^y \right\rangle$ remain zero because of the high symmetry of the system.

\begin{figure}[b]
\centering{
\includegraphics[width=1\linewidth,keepaspectratio,trim=0in 1px 0in 0px, clip]{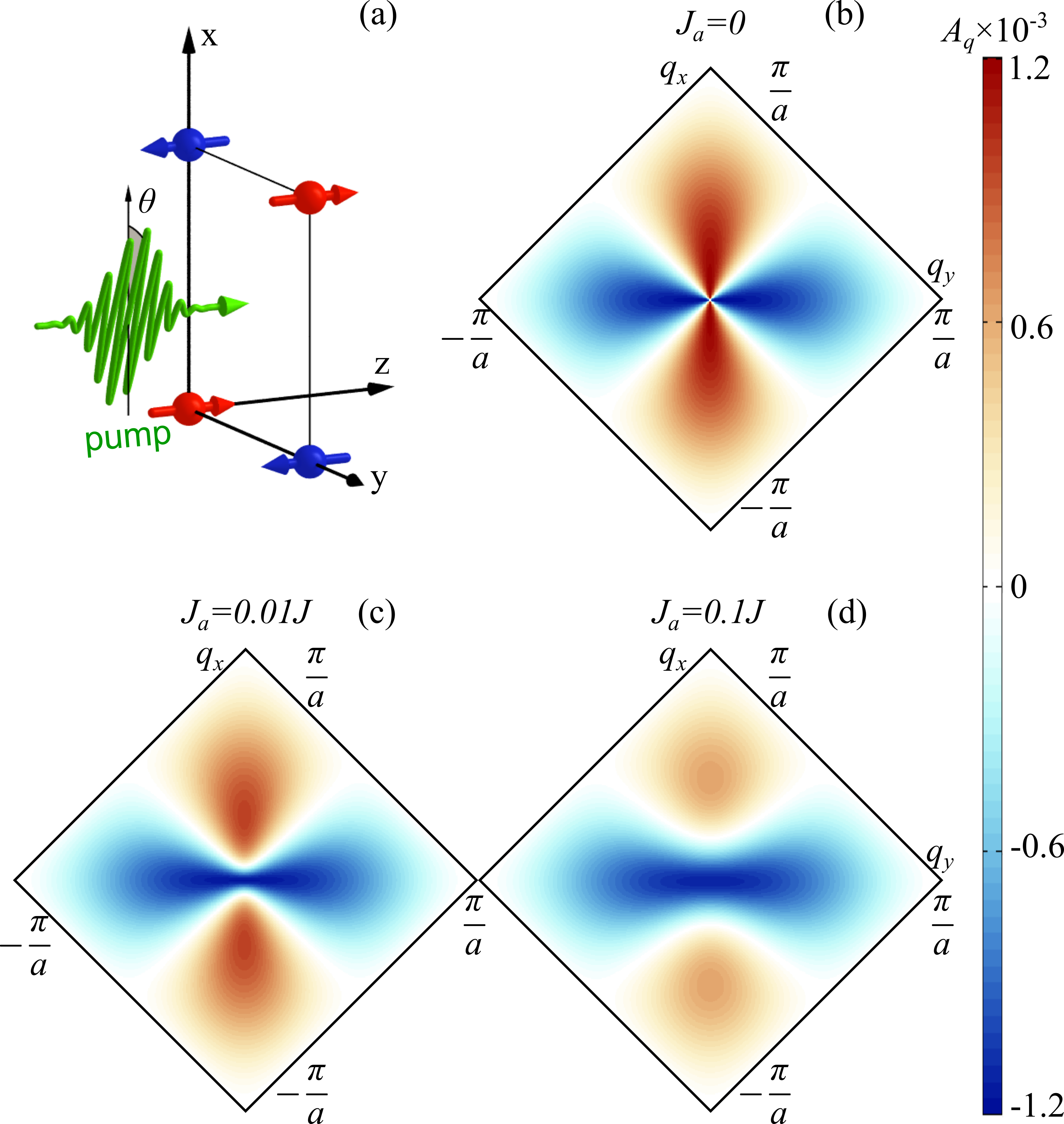}
}
\caption{\label{fig:NeelVec} Illustration of 2D square lattice antiferromagnetic structure (a). Amplitude of oscillations $A_q$ of $z$-projection of spin correlations in the 1-st Brillouin zone(b), without spin anisotropy ($J_a=0$), (c,d) with anisotropy ($J_a=0.01J$, $J_a=0.1J$, respectively). 
Calculations are performed for the same set of parameters as in Fig.\,\ref{fig:SpinCorr}. 
Pump polarization is $\theta=0^\mathrm{o}$.}
\end{figure}

In order to illustrate the origin of this result, we analyzed the simplified case of a square lattice (Fig.\,\ref{fig:NeelVec}(a)). 
In this case, the expression for the changes of $\left\langle L^z(t) \right\rangle$ can be written as
\begin{align}
        \Delta \left\langle L^z(t) \right\rangle&=\sum_q A_q \sin (2 \Omega_q t);\label{eq:NeelVector:Z2D}\\
    A_q&=-\frac{\Delta J  S \zeta \sigma_q}{\hbar} \frac{\gamma_q (\xi_q- \gamma_q)}{1-\gamma_q^2} 4\averaging{\hat{K}^z_q},
\end{align}
where the parameters entering expression for the amplitude $A_q$ reduce to $\gamma_q=\frac{1}{2}\cos(q_x\delta_x)+\frac{1}{2}\cos(q_y\delta_y)$ and $\xi_q - \gamma_q=(e_x^2-\frac{1}{2}) \cos(q_x\delta_x)+(e_y^2-\frac{1}{2}) \cos(q_y\delta_y)$. 
By applying a mirror transformation $R$ along the $(1,1)$ axis we get that $\gamma_{q}$ and $\xi_q - \gamma_q$ transform as 
$R (\gamma_q)=\gamma_{q}$ and $R(\xi_q-\gamma_q) =(e_x^2-\frac{1}{2}) \cos(q_y\delta_y)+(e_y^2-\frac{1}{2}) \cos(q_x\delta_x)$.
As a result, the sum in the right hand side of Eq.~\eqref{eq:NeelVector:Z2D} over the 1st Brillouin zone is equal to zero. 
To visualize this,  Fig.\,\ref{fig:NeelVec}(b) shows $A_q$ evaluated numerically for different wavevectors in the first Brillouin zone.
From the presented data it is evident that the symmetry of the system results in absence of marcoscopic dynamics of $\left\langle L^z\right\rangle$. 
A similar analysis can be applied to the first Brillouin zone of a 3D cubic antiferromagnet, yielding the same result.
In summary, our analysis of the results reveals that the spin dynamics excited by laser-induced perturbation of the exchange interaction in the isotropic cubic crystal, results in THz oscillations stemming from a microscopic characteristic of the magnetic system - spin correlations, while it does not lead to oscillations of the macroscopic order parameter.
In the following section we analyze how the dynamics of spin correlations can be detected in optical experiments.

\subsection{Laser-induced dynamics of dielectric permittivity and probe polarization} \label{sec:Results:Probe}

As a final step in the analysis we evaluate the perturbation of the dielectric permittivity at the central frequency $\omega$ of the probe pulse. 
In accordance with Eq.~\eqref{eq:Epsilon:SpinCorr}, we focus on the evaluation of 
\begin{align}
    \Delta\varepsilon^{\mu \nu}(t)&=\sum_{\delta} \frac{2J\,\Delta\!\left\langle \hat{\mathbf{S}}_{i}(t) \cdot \hat{\mathbf{S}}_{i+\delta}(t) \right\rangle}{\left(U^2-\hbar^2\omega^2\right)} \frac{Q^2 \delta^\mu  \delta^\nu}{\varepsilon_0 a^3} . \label{eq:DeltaEpsilon}
\end{align}
For the cubic structure considered, the dielectric permittivity is diagonal in the equilibrium: $\varepsilon^{xx}=\varepsilon^{yy}=\varepsilon^{zz}=\varepsilon^{0}.$ 
Excitation of the spin correlation results in perturbation of the diagonal tensor components with dominant frequency at two-magnon frequency $2\Omega_{q^*}$
\begin{align}
\varepsilon(t)&= 
            \begin{pmatrix}
            	\varepsilon^{xx}+\Delta\varepsilon^{xx}(t)  & 0  & 0\\
        		0  & \varepsilon^{yy}+\Delta\varepsilon^{yy}(t)  & 0\\
       			0  & 0  &  \varepsilon^{zz}+\Delta\varepsilon^{zz}(t)
    		\end{pmatrix},
    		\label{eq:DeltaEpsilonDiagonal}
\end{align}
with in general $\Delta\varepsilon^{xx}\neq\Delta\varepsilon^{yy}\neq\Delta\varepsilon^{zz}$, \textit{i.e.} spin correlations can induce optical anisotropy in the otherwise optically isotropic antiferromagnet.

\begin{figure}[t]
\centering{
\includegraphics[width=1\linewidth,keepaspectratio,trim=0in 1px 0in 0px, clip]{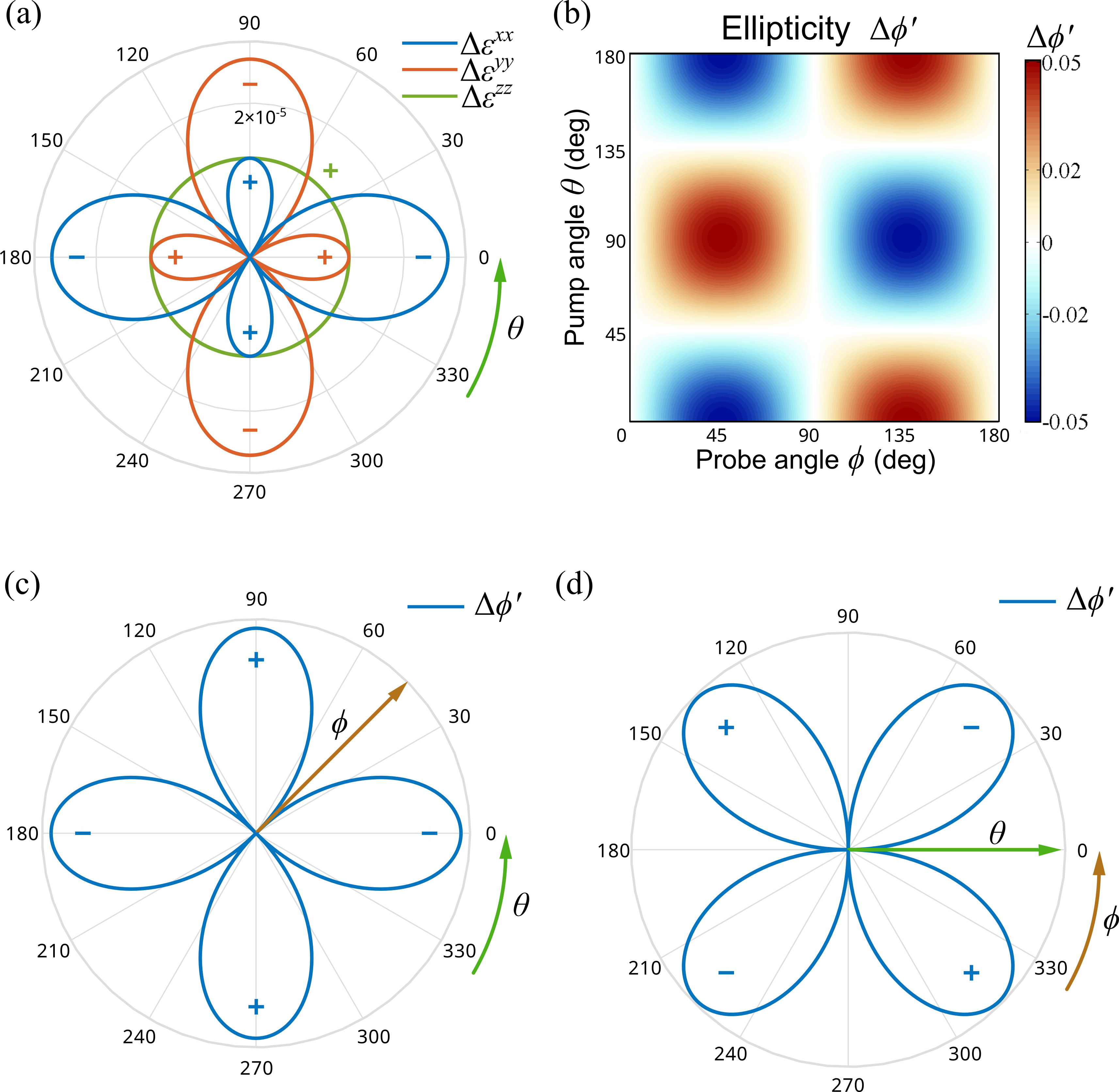}
}
\caption{\label{fig:Ellipticity} Amplitude of laser-induced oscillations of (a) dielectric permittivity tensor components and (b) probe ellipticity $\Delta\phi'(t)$ as a function of pump and probe angles. (c) and (d) present crossections of the ellipticity map, (c) at fixed probe polarization along body diagonal ($\phi=45^\circ$) and (d) at fixed pump polarization along $x$-axis ($\theta=0$). Signs at panels (a,c,d) and colors at panel (b) represent the initial phase of the oscillations.}
\end{figure}

Fig.~\ref{fig:Ellipticity}(a) shows the amplitude and phase of the perturbation of the dielectric tensor components as a function of the pump polarization angle $\theta$ calculated at $t$=9\,fs corresponding to the maximum of the laser-induced changes of spin correlations (see Fig.\,\ref{fig:SpinCorr}(a,c)).
Two important conclusions can be drawn from the presented plot.
First, Fig.\,\ref{fig:Ellipticity}(a) readily reveals that laser-induced dynamics of spin correlations results in optical anisotropy.
For instance, when the pump pulse is polarized at $\theta=0^\mathrm{o}$, $\Delta\varepsilon^{xx}=-2\Delta\varepsilon^{yy}=-2\Delta\varepsilon^{zz}$.
Second, by comparing Fig.~\ref{fig:SpinCorr}(c) and Fig.~\ref{fig:Ellipticity}(a) it is observed that there is a one-to-one correspondence between the amplitude and initial phase of spin correlation for spins along $\mu$-bond and modulation of the dielectric perimittivity tensor component $\Delta\varepsilon^{\mu\mu}$.
Qualitative insight in the emergence of an optically induced anisotropy that appears even in the absence of spin-orbit coupling can also be obtained from the electron hopping in the Hubbard model, as illustrated in Fig.~\ref{fig:Hopping}(c). Again for simplicity we focus on the 2D case, where only a 4x4 lattice is shown with a Neel ground state (Fig.~\ref{fig:Hopping}(c)). 
In accordance with the discussion in Sec.~\ref{sec:Theory:Hubbard}, excitation of the spin system by the pump pulse polarized along $x-$bond yields the mutual hopping of the electrons between the two nearest sites (Fig.~\ref{fig:Hopping}(c)).
This hopping results in the perturbation of the ground state with emergence of the nearest neighbours with parallel spins (Fig.~\ref{fig:Hopping}(d)).
As the hopping between them is blocked, pairs of such sites are excluded from the interaction of the electric field of the probe pulse with the system.
As readily seen in Fig.~\ref{fig:Hopping}(d), the number of such pairs is different along $x-$ and $y-$axes giving rise to optical anisotropy in the $xy-$plane even within the electric-dipole approximation.

Therefore, spin dynamics triggered by the laser-induced perturbation of the exchange interactions manifest itself in the dynamics of  macroscopic observables like the dielectric permittivity. 
In general, the induced modulation of the amplitude and relative sign, \textit{i.e.} the optical anisotropy, gives rise to modulation of intensity of the transmitted or reflected probe pulse and of the probe polarization, both for rotation and for ellipticity. 
For the case of an isotropic transparent antiferromagnet, the modulation of transmission coefficients for the probe electric field components polarized along different bonds will lead to effective rotation of the probe polarization $\Delta\phi(t)$ independent on the sample thickness. 
At the same time, the probe pulse will acquire ellipticity $\Delta\phi'(t)$ increasing with sample thickness. 
For the probe pulse propagating along the $z$-axis (Fig.\,\ref{fig:Setup}(a)) of a sample with thickness $d$ one obtains a simple expression for the induced rotation and ellipticity, as a function of the incoming probe polarization $\phi$ \cite{ImasakaPRB2018}, which reads 
\begin{eqnarray}
    \Delta \phi (t) & = & \frac{\Delta\varepsilon^{xx}(t)-\Delta\varepsilon^{yy}(t)}{2(\sqrt{\varepsilon^{xx}}+1)}\sin(2\phi)\label{eq:DeltaPhi1};\\
    \Delta \phi' (t) & = & \frac{\omega d\left[\Delta \varepsilon^{xx}(t)-\Delta \varepsilon^{yy}(t) \right]}{4  c \sqrt{\varepsilon^{xx}}}  \sin(2 \phi)\label{eq:DeltaPhi},
\end{eqnarray}
where interference effects are neglected, and modulation of the dielectric permittivity is small, $\vert\Delta\varepsilon^{\mu\nu}(t)\vert \ll \varepsilon^{xx}$.

Substituting the solution for spin correlations (Eqs.~(\ref{eq:SpinCorr:DynamicsZ},\ref{eq:SpinCorr:DynamicsXY})) into Eqs.~(\ref{eq:DeltaPhi1},\ref{eq:DeltaPhi}), we numerically calculate the amplitude of oscillations of the probe ellipticity $\Delta\phi'(t)$ for different pump and probe polarization angles $ \mathcal{\theta}$ and $ \phi$, respectively. The result is shown in Fig.\,\ref{fig:Ellipticity}(b,c) where both the pump and the probe propagate along $z$-axis. 
Naturally, the pump polarization dependence of the modulation of the probe ellipticity (Fig.\,\ref{fig:Ellipticity}(b)) resembles that of the dielectric permittivity and spin correlations, and have maxima when pump the is polarized along the bonds. 
The dependence of $\Delta\phi'(t)$ on incoming probe polarization $\phi$ possesses maxima when the probe makes and angle of 45~degree with respect to bond directions.

Figures~\ref{fig:Ellipticity}(b,c) outline the complete experimental strategy of excitation of ultrafast modulation of the exchange energy in an antiferromagnet by a femtosecond laser pulse and detection of subsequent spin dynamics at the two-magnon mode frequency.
Selecting the experimental geometry with pump and probe pulses polarized along and at 45~degree with respect to nearest-neighbor exchange bonds, respectively, yields maximum sensitivity to the spin correlations dynamics.
It is also worth noting that transient probe polarization rotation $\Delta\phi(t)$ and ellipticity $\Delta\phi'(t)$ both scale linearly with the induced optical anisotropy, and possess the same dependence on the incoming pump and probe polarizations. 
However, the coefficients entering expressions (Eqs.~(\ref{eq:DeltaPhi1},\ref{eq:DeltaPhi})) are of the same order of magnitude only for a case of small $d\sim 1\,\mu$m. For thicker samples, the modulation of the probe polarization ellipticity is expected to be more pronounced as compared to the rotation.

\subsection{\label{sec:Results:Anisotropy} Influence of magnetic anisotropy}

As shown in Sec.\,\ref{sec:Theory:DynamicsOfSpinCorr}, in an isotropic cubic antiferromagnet, perturbation of the exchange interaction by ultrashort laser pulses yields dynamics of spin correlations. 
The latter, in turn, manifest itself via modulation of such macroscopic parameter, as a dielectric permittivity tensor. 
At the same time, longitudinal dynamics of the magnetic order parameters, the N\'{e}el vector, is absent. However, in majority of antiferromagnets, there are preferable directions of the spins associated with the presence of magnetocrystalline anisotropy. 
Therefore, it is instructive to examine, if the situation described above holds in anisotropic cubic antiferromagnet as well. 

We consider the uniaxial anisotropy of an easy-axis type, with axis coinciding with the quantization axis $z$. 
Then magnetocrystalline anisotropy can be introduced in the Hamiltonian (Eq.~\eqref{eq:Ham:Heisenberg}) as
\begin{align}
		\hat{H}&=J \sum_{i \delta} \hat{\mathbf{S}}_i \cdot \hat{\mathbf{S}}_{i+\delta}+J_a \sum_{i \delta} \hat{S}^z_i \hat{S}^z_{i+\delta}.\label{eq:Ham:Anisotropic}
\end{align}
We note that, although magnetocrystalline anisotropy is defined for a single magnetic ion, the form of the additional term in Eq.~\eqref{eq:Ham:Anisotropic} yields the same additional contribution to the two-magnon Hamiltonian (Eq.~\eqref{eq:Ham:2M}).
As in the isotropic case, we consider that the laser excitation results in perturbation of the exchange parameter $J$ between the nearest neighbours when the pulse is polarized along the corresponding bond (Eq.~\eqref{eq:Ham:Perturbation:Spin}).
Introduction of magnetic anisotropy results in modification of the parameters $\Omega_q$, $\gamma_q$, and $V_q$ (Eqs.~(\ref{eq:Omega},\ref{eq:Vq})) of the system and excitation as follows:
\begin{align}
		\Omega^*_q&=\frac{z(J+J_a) S}{\hbar}\sqrt{1-\left(\gamma^*_q\right)^2};\\
		\gamma^*_q&=\frac{J}{J+J_a} \gamma_q;\\
        V^*_q&=\frac{\zeta \Delta J S}{\hbar} \frac{\xi_q-\gamma^*_q}{\sqrt{1-\left(\gamma^*_q\right)^2}};\\
        \sigma^*_q&=\sqrt{\pi} \tau_p e^{-(\Omega^*_q)^2 \tau_p^2}.
\end{align}

Changes of parameters $\Omega^*_q$, $\gamma^*_q$, and $V^*_q$ affect results for laser-driven spin correlations. 
However, the general features, such as dependence of the pump polarization $\theta$, remain qualitatively the same.
Furthermore, modulation of dielectric permittivity tensor  related to the spin correlations retains its character, i.e. only modulation of diagonal components of the tensor with different magnitude and sign occurs (Eq.~\eqref{eq:DeltaEpsilonDiagonal}).

The most important difference with the isotropic case occurs when we consider longitudinal dynamics of the N\'eel vector:
\begin{align}
        \Delta \left\langle L^z(t) \right\rangle&=\sum_q A^*_q \sin (2 \Omega^*_q t);\label{eq:NeelVector:ZAnis}\\
    A^*_q&=-\frac{\Delta J  S \zeta \sigma^*_q}{\hbar} \frac{\gamma^*_q (\xi_q- \gamma^*_q)}{1-(\gamma^*_q)^2}\nonumber\\
    &\approx A_q- \frac{J_a}{J+J_a} \frac{\Delta J  S \zeta \sigma_q}{\hbar} \frac{\gamma_q^2 }{1-\gamma_q^2}.\nonumber
\end{align}

Numerical calculations using Eqs.~(\ref{eq:NeelVector},\ref{eq:SpinCorr:DynamicsZ},\ref{eq:SpinCorr:DynamicsXY}) show that, in the anisotropic antiferromagnet, the laser-driven spin correlations result in oscillations of the N\'eel vector at the frequency of the two-magnon mode $2\Omega_{q^*}$.
However, distinct from the situation with spin correlations, the oscillations of the N\'{e}el vector are independent on the pump polarization.
The initial phase of the oscillations of $\Delta \left\langle L^z(t) \right\rangle$ reveals that the system is first driven in the direction with reduced long-range order, with subsequent oscillations around the equilibrium value of $L^z$ which is reduced as compared to the classical value $2S$ due to quantum fluctuations. This is consistent with the behavior revealed for spin correlations (Fig. 3(b)).

Occurrence of the longitudinal dynamics of the N\'eel vector in the cubic crystal with magnetic anisotropy can be illustrated by considering simplified 2D case (Fig.\,\ref{fig:NeelVec}(a)) and using Eq.~\eqref{eq:NeelVector:Z2D} with modified parameters. 
Figs.\,\ref{fig:NeelVec}(c,d) show distribution of the amplitudes $A^*_q$ across the 1st Brillouin zone for two values of $J_a/J$. 
The presence of the magnetic anisotropy results in an additional isotropic contribution to $A^*_q$, breaking of the mirror symmetry with respect to the $(1,1)$ axis. 
This additional contribution to $A^*_q$ is found to be independent on the pump polarization leading to the absence of the dependence of oscillations of N\'eel vector on $\theta$.

It is important to stress, that the obtained result on the dynamics of N\'eel vector is independent on the orientation of the anisotropy axis. 
Thus, a pump pulse propagating along arbitrary direction in the anisotropic cubic antiferromagnet would excite dynamics of spin correlations governed by the angle of the electric field of the pump pulse makes with respect to the different bonds, and the pump-independent longitudinal dynamics of the N\'eel vector. 
Both types of excited spin dynamics can be probed optically by choosing a proper propagation direction and polarization of the probe pulse.
Dynamics of the N\'eel vector changes the optical properties via magnetic linear birefringence - an effect of spin-orbital origin not accounted for by our microscopic model.
Therefore, we analyze the effect of laser-driven oscillations of $\left\langle L^z(t) \right\rangle$ within the phenomenological approach based on space-time symmetry \cite{SmolenskiUsp1975}.

In the considered geometry (Fig.\,\ref{fig:Setup}(a)) the time-dependent N\'eel vector induces additional difference between dielectric tensor components $\varepsilon^{xx(yy)}(t)-\varepsilon^{zz}(t)=b\left(\left\langle L^z(t) \right\rangle\right)^2\approx b\left\langle L^z \right\rangle^2+2b\left\langle L^z \right\rangle\left\langle \Delta L^z(t) \right\rangle$, where $b$ is the magneto-optical coefficient. 
Thus, oscillations of the N\'eel vector could not be sensed by the pulse propagating along $z$-axis.
Since in experiments pump and probe pulses often propagate nearly collinearly to each other, we  consider here an additional case with the pump and the probe propagating along the $y$-axis in the layout shown in Fig.\,\ref{fig:Setup}(a). 
In this case 
\begin{align}
    \Delta \phi' (t) & =  \frac{\omega d}{4  c } \left[ \frac{\Delta\varepsilon^{xx}(t)}{\sqrt{\varepsilon^{xx}+b\left\langle L^z(t) \right\rangle^2}}-\frac{\Delta \varepsilon^{zz}(t)}{\sqrt{\varepsilon^{xx}}}\right.\nonumber\\
   	&\quad 
   	+ \left.\frac{2b\left\langle L^z \right\rangle}{\sqrt{\varepsilon^{xx}+b\left\langle L^z(t) \right\rangle^2}}\left\langle \Delta L^z(t) \right\rangle \right] \sin(2 \phi)\label{eq:DeltaPhiL},
\end{align}
where $\Delta\varepsilon^{xx}$ and $\Delta\varepsilon^{zz}$ describe modulation due to spin correlations directly and correspond to $\Delta\varepsilon^{xx}$ and $\Delta\varepsilon^{yy}$, respectively, in Fig.\,\ref{fig:Ellipticity}(a).
Thus, oscillations of N\'eel vector bring additional contribution to the dynamic ellipticity of the probe pulse. 
However, this contribution is independent on the pump polarization $\theta$ and thus can be distinguished from the contribution from $\Delta\varepsilon^{xx}$ and $\Delta\varepsilon^{zz}$.

\section{Discussion}\label{sec:Discussion}

Before discussing the results obtained in light of experimental findings reported so far, we note that our result can be easily generalized to other antiferromagnets, for example those featuring non-cubic crystal structures. The main requirement for a magnetic structure to support the results obtained in our analysis, is the presence of inversion symmetry, which allows us to inverse a sign of magnon wavevectors. 
Another condition is the transparency of the crystal for both pump and probe pulses, which keeps changes of double occupancy negligible.
Thus we argue that the results obtained in our analysis are applicable for dielectric weakly-absorbing materials with crystal structures possessing an inversion center, having one leading contribution to the isotropic exchange interaction, and without anisotropic contributions to exchange energy, such as Dzyaloshinskii–Moriya interaction.
Materials in which laser-driven two-magnon modes were demonstrated experimentally, transition metal fluorides, do satisfy such requirements to a certain degree.

\subsection{Comparison with experimental data:\\quantitative analysis}\label{sec:Experiment_values}

We start from estimating magnitude of changes $\Delta J$ expected for the pump fluences used in a typical experiment with transparent dielectrics.
We use KNiF$_3$ as a model material, since its magnetic and crystal structure can be readily described by the structure shown in Fig.\,\ref{fig:Setup}(a), and all relevant material parameters and an extended set of experimental data on laser-driven two-magnon mode on this material is available.

KNiF$_3$ crystallizes into cubic structure, while antiferromagnetic ordering is of G-type and is characterized by a single exchange constant.
The N\'eel vector is aligned with one of cubic axes $\langle100\rangle$, and the magnetic anisotropy is weak.
The material is optically transparent in near-infrared and visible range, and optically isotropic in equilibrium.
Material parameters $J$, $S$, $U$, $a$, and $\varepsilon^{xx}$ required for calculations are listed in Table\,\ref{table:parameters}.
Experiments on excitation of two-magnon mode by a laser pulse were reported in \cite{BossiniNatComm2016,Bossini2019}.
Parameters of experiments, such as laser fluence $F$, pump pulse duration $\tau_p$, pump and probe polarizations, and the amplitude of the oscillations of the probe polarization $\Delta\phi$ being the outcome of experiments are presented in Table\,\ref{table:results}.
Experiments were performed with photon energies of pump and probe pulses being $\hbar\omega_p$=2.2\,eV and $\hbar\omega$=1.3\,eV.
In this spectral range the dispersion of the equilibrium dielectric permittivity of KNiF$_3$ can be neglected and a single value of $\varepsilon^{xx}=\varepsilon^{yy}=\varepsilon^{zz}$ can be used.
Note that the experimental value of $\Delta\phi$ is provided per 1$\mu$m in Table\,\ref{table:results}, while in experiments the sample thickness was of $d=340$\,$\mu$m.

Since KNiF$_3$ possesses low magnetic anisotropy, we argue that the dominating contribution to the measured modulation of the probe polarization originates from the induced optical birefringence, Eq.~\eqref{eq:DeltaEpsilon}, and not from oscillations of the N\'eel vector and related magnetic linear birefringence.
For calculating the impact of the pump pulse on the exchange interaction $J$ we estimate the electric field amplitude of the pump pulse $E_0$ as
\begin{align}
    E_0^2=\frac{(1-R)F}{\sqrt{\ln2}\varepsilon_0 \varepsilon^{xx} c \tau_p},
\end{align}
where $c$ is the speed of light in vacuum and $R=0.035$ is the reflectivity coefficient.

Using Eq.~\eqref{eq:DeltaJ:Gauss}, we obtain a relative perturbation of the exchange energy of $\Delta J/J = 0.008$ for $\theta=0,\pi/2$.
In Table\,\ref{table:results} we report the corresponding modulation of optical parameters of the medium at the probe pulse photon energy and resulting modulation of the probe ellipticity $\Delta\phi'$ for probe polarized at $\phi=\pi/4$. 
For the sake of generalization, we report $\Delta\phi'$ per 1~$\mu$m material thickness.  

\begin{table}
    \centering
\caption{Experimental parameters and results of observation of laser-induced two-magnon mode in KNiF$_3$, and the results of our calculations.}
\begin{tabular}{c|c|c|c|c|c|c|c}
  \hline
  \multicolumn{8}{c}{theory {[this work]}}\\
  \hline
   $E_0$ & $\tau_p$ & $\theta$ & $\Delta J/J$ & $\Delta\varepsilon^{xx}$ & $\Delta n$  & $\phi$  & $\Delta\phi'$ \\
   V/cm & fs & rad & & & & rad & mrad/$\mu$m\\
   \hline
   \multirow{2}{*}{1.9 $\cdot 10^9$} & \multirow{2}{*}{5}  & \multirow{2}{*}{0}  & \multirow{2}{*}{0.008}  & \multirow{2}{*}{2.8 $\cdot 10^{-5}$} & \multirow{2}{*}{9.6 $\cdot 10^{-6}$}  & $\pi/4$   & 0.05    \\
  &  &   &   &   &   & $\pi/20$  & 0.016    \\
  \hline
  \multicolumn{8}{c}{experiment with KNiF$_3$ \cite{Bossini2019}} \\
  \hline
  $F$ & $\tau_p$ & $\theta$ & \multicolumn{3}{c|}{} & $\phi$  & $\Delta\phi$ \\
  mJ/cm$^2$  & fs & rad & \multicolumn{3}{c|}{} & rad & mrad/$\mu$m  \\
  \hline
  8.6 & 5 & 0 & \multicolumn{3}{c|}{} &$<\pi/20$  & 0.017 \\
   \hline
    \hline
\end{tabular}
    \label{table:results}
\end{table}

In \cite{BossiniNatComm2016}, the detection in experiments has been realized using probe pulses polarized at $\phi<\pi/20$ which decreases the detected amplitude of polarization modulation by a factor of $\sim$0.3 as compared to that found for $\phi=\pi/4$ (Eq.~\eqref{eq:DeltaPhi}).
Indeed, our calculations for $\phi=\pi/20$ yield the value of $\Delta\phi'=0.016$, which is in a good agreement with the experimental results (Table\,\ref{table:results}).
Thus, we argue that the outcome of our theoretical model quantitatively describes the results of experiments with KNiF$_3$. In Sec.\,\ref{sec:Experiment_symmetry} we substantiate it further by analysing polarization dependencies of the probe polarization modulation in KNiF$_3$.

In order to put the obtained value of $\Delta J/J$ and related modulation of optical birefringence in perspective and to justify further if these values are realistic, we compare $\Delta J/J$ and $\Delta n$ to the changes of these values in KNiF$_3$ under strain $\Delta a/a$ available in literature.
Applying relation $J\sim a^{-12}$ \cite{deJongh_PhysB1975}, we estimate that that for KNiF$_3$, $\Delta J/J=0.01$ effectively corresponds to compressive strain $\Delta a/a\approx10^{-3}$, which is a large, but experimentally accessible value.

Having established the equivalent strain required to achieve a change of the exchange interaction by 1\%, we can now compare the induced optical birefringence obtained in our work with experimental values of strain-induced birefringence reported for KNiF$_3$.
According to \cite{Ferre_JPC1983}, the optical birefringence induced by strain of $10^{-3}$ applied along $\langle001\rangle$ is $\vert\Delta n\vert\sim10^{-4}$ in paramagnetic phase and $\vert\Delta n\vert<10^{-4}$ below $T_N$. 
The calculated optical birefringence $\vert\Delta n\vert\approx1\cdot10^{-5}$ (Table\,\ref{table:results}) is in reasonable agreement with this value, supporting validity of calculations of optical birefrinence induced by changes of $\Delta J$. 

Thus, by comparing results of our calculations with experimental data on laser-induced two-magnon mode and to the static data on strain-induced modulation of exchange parameter and optical birefringence, we obtained a realistic estimation of the impact of the laser pulse on $J$ reaching $\approx$1\% for laser fluences typically used for pump-probe experiments.

\subsection{Comparison with experimental data:\\symmetry analysis}\label{sec:Experiment_symmetry}

\subsubsection{Antiferromagnet with cubic lattice KNiF$_3$}

The comprehensive experimental study of the laser-driven two-magnon mode in KNiF$_3$ \cite{BossiniNatComm2016,Bossini2019} also provides the basis for analysing applicability of our model in terms of a symmetry of the effect.
Several key observations were reported on the excitation and detection of two-magnon mode in this material.
First, it has been reported that the oscillations of the probe polarization at the frequency of the two-magnon mode are dependent on the pump polarization.
For efficient excitation, pump pulses were polarized along $\langle100\rangle$, i.e. along $\boldsymbol{\delta}$ for this material.
This is in excellent agreement with calculations based on our models as  illustrated in Fig.\,\ref{fig:SpinCorr}(c).
Second, the initial phase of the detected oscillations of the probe polarization was changing by $\pi$ upon rotation of the pump polarization from one axis to another.
Again, such pronounced pump-polarization dependence of the detected probe polarization dynamics suggests that the macroscopic parameter responsible for the measured signal is the induced optical birefringence (Eq.~\eqref{eq:DeltaEpsilon}), and not magnetic linear birefringence emerging due to oscillations of $\langle L^z(t)\rangle$. 
The latter should be absent due to a weak anisotropy of the system.
Thus, the experimental results reported in \cite{Bossini2019} support one of the main conclusions of our analysis predicting that the induced optical birefringence is the leading macroscopic manifestation of the laser-driven two-magnon mode.

It is important to point out that the authors of \cite{BossiniNatComm2016,Bossini2019} have attributed the oscillations of the probe polarization to the oscillations of $\langle L^z(t)\rangle$, which have led to the necessity to introduce additional symmetry lowering of the system into the model.
Indeed, the induced oscillations of $\langle L^z(t)\rangle$ are pump polarization independent in both our analysis and in the analysis reported in \cite{Bossini2019}.
Therefore, in order to explain the pronounced pump polarization dependence of the signal, in \cite{Bossini2019} it was suggested that additional magnetoelastic strain or spin-orbit effects contribute to different signs of $\Delta J$ when pump is polarized along and perpendicularly to the equilibrium direction of $\mathbf{L}$.

\subsubsection{Antiferromagnets with tetragonal lattice MnF$_2$, FeF$_2$, and CoF$_2$}

Applicability of our analysis to antiferromagnets with non-cubic lattice structures can be demonstrated by considering the results on excitation and detection of the two-magnon mode in transition metal difuorides MnF$_2$, FeF$_2$ \cite{ZhaoPRL2004,ZhaoPRB2006}, and CoF$_2$ \cite{Formisano_JPCM2022} possessing tetragonal rutile symmetry.
The dominating exchange interaction in these antiferromagnets is between spins belonging to different sublattices with $\boldsymbol{\delta}\|
\langle 111 \rangle$, while the N\'eel vector is along the [001] axis.
Depending on the transition metal ion, materials possess stronger (CoF$_2$) or weaker (MnF$_2$) magnetic anisotropy. All these materials are optically anisotropic with [001] being the optical axis.

In experiments reported in \cite{ZhaoPRL2004,ZhaoPRB2006} pump and probe pulses were propagating along the [001] axis.
Thus, equilibrium optical anisotropy in the (001) plane is absent, and the analysis  of optical response linked to laser-driven dynamics of spin correlations (Eqs.~(\ref{eq:DeltaEpsilonDiagonal}-\ref{eq:DeltaPhi})) holds.
The pump polarization was along [110] direction, which coincides with the projection of $\boldsymbol{\delta}$ on the (001) plane.
While pump polarization dependence of the detected signal was not reported, the chosen pump polarization agrees with the outcome of our analysis where the electric field of the pump pulse should have a projection on $\boldsymbol{\delta}$ in order to affect the hopping $t_{ij}$ and, thus, yield nonzero $\Delta J$.
Differential transmission for the probe pulses polarized along and perpendicular to the pump pulse were measured, which corresponds to the rotation of the probe polarization described by Eq.~\eqref{eq:DeltaPhi1}.
In agreement with our analysis, the probe was initially polarized at 45\,degree with respect to the pump pulses, thus being the most sensitive to induced optical anisotropy at the two-magnon frequency. 
Longitudinal dynamics of the N\'eel vector aligned along [001] axis cannot be optically probed in this geometry.
We note, that the choice of the pump and probe polarizations in \cite{ZhaoPRL2004,ZhaoPRB2006} was made based on the symmetry of the two-magnon mode and the corresponding Raman tensor.

In \cite{Formisano_JPCM2022} the pump and the probe pulses were propagating in CoF$_2$ crystal along [010] axis, and the equilibrium optical birefringence strongly affected the polarization of pump and probe pulses, making direct comparision of the experimental results with our theory intricate.
However, this geometry is of particular interest to us, since modulation of the probe parameters due to both changes of optical birefringence $\Delta\varepsilon^{\mu\nu}(t)$ and longitudinal dynamics of $\langle L^z(t)\rangle$ can be observed.
The authors reported that the detection of the laser-driven two-magnon mode realized by measuring transient probe rotation was possible with the probe pulses polarized along [001] axis.
When the probe pulse was polarized at 45\,degree, the detected signal was vanishing. 
On the one hand, such observation contradicts the scenario where the longitudinal dynamics of the N\'eel vector is detected via magnetic linear birefringence. 
Indeed, in order to observe such contribution, one need to employ probe pulses polarized at 45\,deg to N\'eel vector, i.e. to [001] axis. 
On the other hand, probe pulses polarized along [100] or [001] axes, as was used in the reported experiments, would be sensitive to additional optical birefringence induced between $\langle101\rangle$ directions in the (010) plane.
Thus, experimental results on laser-driven two-magnon mode in CoF$_2$ \cite{Formisano_JPCM2022} suggest that the transient optical birefringence remains dominating macroscopic manifestation of the excited dynamics of spin correlations even in material with strong magnetic anisotropy. 
We note that no pronounced pump polarization dependence of the time-resolved signal at $2\Omega_{q^*}$ could be extracted from the data, which is a result of the static optical  birefringence, as well as of a poor signal-to noise-ratio.
However, somewhat stronger signals were observed with pump pulses polarized along $[101]$, i.e. when there is a maximal projection of the pump electric field on $\boldsymbol{\delta}\|\langle111\rangle$.
Thus, overall, all results reported for the transition metal fluorides are well described by our model, even though the crystal structure of these materials is different from the cubic lattice.

\section{Conclusions}

We have developed a full theoretical description of the excitation, dynamics, and the detection of spin correlations by femtosecond laser pulses based on a minimal model of a transparent cubic antiferromagnet. 
Despite the simplicity of our model, it features observable responses in prototype strongly correlated systems, which agree with existing experimental reports.
%The unified approach based on Heisenberg and Hubbard models
A single theoretical framework which allows us to describe that the impulsive perturbation of the exchange interaction by an electric field of the femtosecond laser pulse triggers oscillations of spin correlations, with initial phases and amplitudes different for the bonds along and perpendicular to the electric field of the pulse.
Since the electric field of light affects the interaction between the nearest neighbors, dynamics of spin correlations is dominated by the frequency of the two-magnon mode. 
The oscillations can be understood as a periodic shift of spin correlations along one of the axes closer to the values of local singlet or N\'eel states, and shift of pairs along other two axes in the opposite directions, while the system as a whole remains close to the long-range ordered N\'eel state.
By varying the pump polarization, we obtain the complete pump polarization dependence of spin correlation dynamics and demonstrate that it is the orientation of the electric field of the pulse with respect to the bonds between the nearest neighbors that governs this dependence.

We further show that in cubic antiferromagnets without magnetic anisotropy, perturbation of exchange interation does not yield dynamics of the N\'eel vector, and, thus, dynamics of spin correlations cannot be explained in terms of classical macroscopic parameters such as ferromagnetic or the Néel vector.
This is in contrast to the previously suggested scenario that oscillations of the N\'eel vector are the macroscopic manifestation of the laser-driven excitation of the two-magnon mode \cite{Bossini2019,BossiniNatComm2016}.
We show that inclusion of strong uniaxial anisotropy in the model leads to the emergence of laser-induced dynamics of the N\'eel vector, because the spin-orbit interaction, which determines magnetic anisotropy, breaks independence between spin and lattice subsystems. 
This contribution appears, however, only in next to leading order, determined by the smallness of spin-orbit with respect to exchange interactions.
Moreover, in cubic crystal it does not discriminate between perturbation of exchange along different bonds, and is therefore insensitive to polarization of the excitation pulse.

The developed model also reveals that the macroscopic manifestation of the laser-driven exchange perturbation and dynamics of spin correlations in cubic antiferromagnets is the anisotropic modulation of the diagonal components of the dielectric permittivity tensor of the electric-dipole type. 
As a result, dynamics of spin correlations can be detected in the probe polarization ellipticity or rotation. 
Again, this effect is governed by the crystal symmetry and does not depend on the orientation of the antiferromagnetic vector.
The latter dependence emerges only when anisotropy is included and the oscillation of the N\'eel vector is excited leading to modulation of magnetic linear birefringence. 

To substantiate the conclusions drawn on the basis of the developed model, we examined the symmetry and magnitude of the laser-induced dynamics of the two-magnon mode reported in literature, and compared those to theoretical predictions.
We compared pump and probe polarization dependencies to experimental results in cubic (KNiF$_3$) and tetragonal (MnF$_2$, FeF$_2$, CoF$_2$) antiferromagnets.
The experimental results for KNiF$_3$ reported in \cite{Bossini2019} show good agreement with our theory, both in terms of magnitude and symmetry of the effect. Although the theory is developed for cubic crystal, the results obtained can be easily generalized to the case of centrosymmetric crystals. This is of great relevance for studies on transition metal fluorides such as FeF$_2$ and CoF$_2$ \cite{ZhaoPRL2004,ZhaoPRB2006,Formisano_JPCM2022} for which no experimental polarization analysis has been reported so far. 
We hope that our theory stimulates further experiments on the ultrafast dynamics of spin correlations and facilitates the development of intuitive models for the dynamics magnetism at the shortest length and time scale.

\begin{acknowledgments}
Authors thank R.M. Dubrovin, A.V. Kimel, R.V. Pisarev, F. Formisano, M. D. Bouman and G. Fabiani for insightful discussions.
The work of A.E.F. was partially supported by BASIS Foundation (grant No.~20-1-5-95-1), EU COST Action CA17123 Magnetofon under Short term scientific mission program (2020), as well as Russian Science Foundation (grant no.~22-72-00039). 
J.H.M. acknowledges funding the Shell-NWO/FOM-initiative “Computational sciences for energy research” of Shell and Chemical Sciences, Earth and Life Sciences, Physical Sciences, FOM, and STW, as well as funding from the European Research Council under ERC Grant Agreement No. 856538 (3D-MAGiC) and the Horizon Europe project no. 101070290 (NIMFEIA).
\end{acknowledgments}

\appendix
\section{Polarization and spin operators}\label{App:PolToSpin}
We define new hopping operators as
\begin{align}
		 \tilde{T}^\mu&= \sum_{i,\delta} {\delta}^\mu \tau_{i,i+\delta} \hat{c}^{\dagger}_i \hat{c}_{i+\delta};\\
		 \tilde{T}^{\mu \nu}&=\sum_{i,\delta} {\delta}^\mu {\delta}^\nu\tau_{i,i+\delta} \hat{c}^{\dagger}_i \hat{c}_{i+\delta}. \nonumber
\end{align}
These definitions allows us to use Schrieffer-Wolff transformation for commutators with polarization operators 
\begin{align}
		\langle \psi_n| [\hat{P}^\mu;\hat{H}_{U}] |\psi_m\rangle &=-\frac{\lambda Q}{a^3}\langle \psi_n|  \tilde{T}^\mu |\psi_m\rangle \\
        		&=-\frac{\lambda Q}{a^3}\langle \phi_n|e^{i \lambda\hat{S}^{(1)}} \tilde{T}^\mu 
                	e^{-i \lambda\hat{S}^{(1)}}|\phi_m\rangle \nonumber\\
                &=-\frac{\lambda Q}{a^3}\langle \phi_n|\tilde{T}^\mu |\phi_m\rangle \nonumber\\
                &\quad-\frac{\lambda^2 Q}{U a^3} \langle \phi_n|\left(\hat{T}_- \tilde{T}^\mu_+ +\tilde{T}^\mu_- \hat{T}_+ \right)|\phi_m\rangle, \nonumber
\end{align}
where $|\phi_m\rangle = e^{i \lambda\hat{S}^{(1)}}|\psi_m\rangle $.
As far as electronic transition is virtual, double occupancy for each cite is equal to zero. Taking this into account, we keep only the contributions quadratic in $\lambda$.  Zero double occupancy ground state also leads to $\langle{\hat{T}_+ \hat{T}_-}\rangle =0$. The same approach can be applied to another commutator:

\begin{align}
    &\langle \psi_n| [\hat{P}^\mu;[\hat{P}^\nu;\hat{H}_{U}]] |\psi_m\rangle\\ 
        &\quad= -\frac{\lambda^2 Q^2}{U a^6} \langle \phi_n|\left(T_- \tilde{T}^{\mu \nu}_+ +\tilde{T}^{\mu \nu}_- T_+ \right)|\phi_m\rangle \nonumber\\
        &\quad= -\frac{2\lambda^2 Q^2}{U a^6} \langle \phi_n|\tilde{T}^{\mu }_-\tilde{T}^{\nu}_+  |\phi_m\rangle,\nonumber
\end{align}
where $\hat{T}_- \tilde{T}^{\mu \nu}_+ +\tilde{T}^{\mu \nu}_- \hat{T}_+ = 2 \tilde{T}^{\mu }_-\tilde{T}^{\nu}_+$ due to the definitions of $\tilde{T}^{\mu \nu}$ and $\tilde{T}^{\mu}$.
Then Eq.~\eqref{eq:varepsilon_W_comm} in the main text can be written in terms of new hopping operators whose product is proportional to the spin correlation:
\begin{align}
\varepsilon^{\mu \nu}(\omega) &=\varepsilon^0 I^{\mu \nu}  + \frac{Q^2}{\varepsilon^0 a^3}\frac{2 \langle \tilde{T}^{\mu }_-\tilde{T}^{\nu}_+\rangle}{U \hbar^2 \omega^2}  \\
        &\quad- \frac{Q^2}{\varepsilon^0 a^3}\frac{1}{\hbar^2 \omega^2} \frac{\langle \tilde{T}^{\mu }_-\tilde{T}^{\nu}_+ \rangle} {\hbar\omega+U}+ \frac{Q^2}{\varepsilon^0 a^3}\frac{1}{\hbar^2 \omega^2} \frac{\langle \tilde{T}^{\mu }_-\tilde{T}^{\nu}_+ \rangle} {\hbar\omega-U} \nonumber\\          
        &=\varepsilon^0 I^{\mu \nu} + \sum_{\delta} \frac{4 |\tau_{i j}|^2\left(\averaging{\hat{\mathbf{S}}_{i} \cdot \hat{\mathbf{S}}_{i+\delta}}-\frac{1}{4} \right)}{ U \left(U^2-\hbar^2\omega^2\right)}
    		\frac{Q^2  \delta^\mu \delta^\nu}{\varepsilon^0 a^3}. \nonumber           
\end{align}

\section{Two-magnon operators dynamics}\label{App:Kdynamics}
In this Appendix, we show how to derive two-magnon dynamics in case of arbitrary perturbation profile. As we discussed in Sec. \ref{sec:Theory:Perturbation}, perturbation can be expressed as
\begin{align}
     \Delta \hat{H}_{2M} &=f(t) \sum_{q} \hbar V_q (\hat{K}^+_q+\hat{K}^-_q)\\
     &=2 f(t) \sum_{q} \hbar V_q \hat{K}^x_q=f(t) \Delta \hat{H}_{2M,0}, \nonumber
\end{align}
where
\begin{align}
        \left[\hat{K}^x_q,\hat{K}^y_q\right] &= -i \hat{K}^z_q;\\
        \left[\hat{K}^y_q,\hat{K}^z_q\right] &= i \hat{K}^x_q ;\nonumber\\
        \left[\hat{K}^z_q,\hat{K}^x_q\right] &= i \hat{K}^z_q. \nonumber
\end{align}

We define $\hat{A}_q$ as one of $\hat{K}_q$ operators and write Kubo formula for it
\begin{align}
    \averaging{\hat{A}_q (t)}&=\averaging{\hat{A}_{q,0}} - \frac{i}{\hbar}\int_{-\infty}^t \averaging{[\hat{A} _q (t),\Delta \hat{H}_{2M}(\tau)]} d\tau\\ 
    	&=\averaging{\hat{A}_{q,0} } + 2 V_q \int_{-\infty}^{\infty} f(\tau) G(\hat{A}_q, \hat{K}_q^x|t-\tau ) d\tau\nonumber\\
    	&=\averaging{\hat{A}_{q,0}} + 2 V_q \int_{-\infty}^{\infty} f(\omega) G(\hat{A}_q, \hat{K}_q^x|\omega ) e^{i \omega t} d\omega. \nonumber
\end{align}

Green functions can be found from equations of motion for $G(A_q, K_q^x|\omega)$ and $G(A_q, K_q^y|\omega)$
\begin{align}
	\omega G(\hat{A}_q, \hat{K}_q^x|\omega)&=\frac{1}{\sqrt{2 \pi}} \averaging{[\hat{A}_q, \hat{K}_q^x]}-2i \Omega_q  G(\hat{A}_q, \hat{K}_q^y|\omega)\\
    \omega G(\hat{A}_q, \hat{K}_q^y|\omega)&=\frac{1}{\sqrt{2 \pi}} \averaging{[\hat{A}_q, \hat{K}_q^y]}+2i \Omega_q  G(\hat{A}_q, \hat{K}_q^x|\omega)
\end{align}
Next, we solve this system and obtain the Green function for $\hat{A}_q=\hat{K}^x_q$
\begin{align}
	G(K^x_q, K_q^x|\omega)&=-\frac{1}{\sqrt{2 \pi}} \frac{2 \Omega_q \averaging{K_q^z}}{\omega^2-(2\Omega_q)^2} 
\end{align}
In case of Gaussian pulse $f(t)=e^{-t^2/\tau_p^2}$ we obtain
\begin{align}
 	\averaging{K^x_q(t)}&=-\frac{2 V_q }{2i} \left(F(-2i\Omega_q)- F(2i\Omega_q)\right) \averaging{K^z_{q,0}},
\end{align}
where $F(p)=\frac{\sqrt{\pi}}{2}\sigma_p e^{p t} \erfc\left(-\frac{t}{\tau_p}+\frac{p \tau_p}{2}\right)$, and $\sigma_p=\tau_p e^{p^2 \tau_p^2 / 4}$. For $\tau_p\ll \Omega_q^{-1}$ we can replace erfc with the Heaviside function $\Theta(t)$
\begin{align}
    \averaging{\hat{K}^x_q(t)}&=-4 V_q \sqrt{\pi} \tau_p e^{-\Omega_q^2 \tau_p^2} \left\langle \hat{K}^z_{q,0}\right\rangle \sin{(2 \Omega_q t)}\Theta(t).
\end{align}
\bibliography{References}

\end{document}